\documentclass{article}

\usepackage[english]{babel}

\usepackage[a4paper,top=2cm,bottom=2cm,left=3cm,right=3cm,marginparwidth=1.75cm]{geometry}

\usepackage[backend=bibtex, style=numeric, sorting=none]{biblatex}
\usepackage{amsmath}
\usepackage{graphicx}
\usepackage[colorlinks=true, allcolors=blue]{hyperref}
\usepackage{graphicx}
\usepackage{float}
\usepackage{titlesec}
\usepackage{multirow}
\usepackage{longtable}
\usepackage{lscape}
\usepackage[width=\textwidth]{caption}
\usepackage{changepage}
\usepackage{authblk}
\usepackage{hyperref}
\usepackage{multicol}
\usepackage{wrapfig}
\usepackage{csquotes}

\providecommand{\keywords}[1]
{
  \small	
  \textbf{\textit{Keywords---}} #1
}

\title{PlayMolecule pKAce: Small Molecule Protonation through Equivariant Neural Networks}
\author[a,b]{Nikolai Schapin}
\author[a]{Maciej Majewski}
\author[a]{Mariona Torrens-Fontanals}
\author[a,b,c]{Gianni De Fabritiis\thanks{corresponding author, e-mail address: g.defabritiis@acellera.com}}
\affil[a]{Acellera Labs, C/ Doctor Trueta 183, 08005 Barcelona, Spain}
\affil[b]{Computational Science Laboratory, Universitat Pompeu Fabra, PRBB, C/ Doctor Aiguader 88, 08003 Barcelona, Spain}
\affil[c]{Institució Catalana de Recerca i Estudis Avançats (ICREA), Passeig Lluís Companys 23, 08010 Barcelona, Spain}

\begin{document}
\maketitle

\begin{abstract}
Small molecule protonation is an important part of the preparation of small molecules for many types of computational chemistry protocols. For this, a correct estimation of the pKa values of the protonation sites of molecules is required. In this work, we present pKAce, a new web application for the prediction of micro-pKa values of the molecules' protonation sites. We adapt the state-of-the-art, equivariant, TensorNet model originally developed for quantum mechanics energy and force predictions to the prediction of micro-pKa values. We show that an adapted version of this model can achieve state-of-the-art performance comparable with established models while trained on just a fraction of their training data. 
\end{abstract}


\begin{multicols}{2}
\section{Introduction} 
Proper molecule preparation is a crucial step in many computational chemistry protocols. The correct assignment of the protonation state is one of the key tasks, as it widely affects molecular behavior such as solubility, permeability, metabolism, affinity, or excretion kinetics \cite{prot_state_importance}. However, because of chemical diversity, accurate pKa estimation is challenging and experimental determination is infeasible when dealing with large molecular libraries due to being too cost and time expensive. 

Machine learning (ML) methods have proved their utility in many applications in cheminformatics \cite{ml_dd_overview}. They are successfully deployed for binding affinity predictions \cite{kdeep,bindscope,GGL-Score}, physicochemical properties like solubility \cite{sol_ml} and lipophilicity \cite{lipo_ml}, toxicology \cite{ggl_tox} and ADME properties \cite{adme1,admet2,cardio_tox} predictions and in generative chemistry to propose new molecules obeying set chemical criteria \cite{reinvent}. Also in pKa prediction, various models exist, both simple 2D models \cite{ml_meets_pka,pka_rooted_top_fp} as well as more complex graph neural networks \cite{multi-instance_pka,molgpka,epik}. While they have good performances on public benchmark datasets they can show poor performance for specific molecules, functional groups, or scaffolds, mainly due to a restricted design that allows only certain sets of molecules, such as is the case for tree-based models that have difficulties with multiprotic molecules \cite{ml_meets_pka}, or insufficient high-quality data \cite{challenges_pka}. The models also are often specific to the type of molecular structures that they accept and not all models support multiprotic structures \cite{ml_meets_pka}.

\begin{figure*}[!hb]
\centering
     \includegraphics[width=\textwidth]{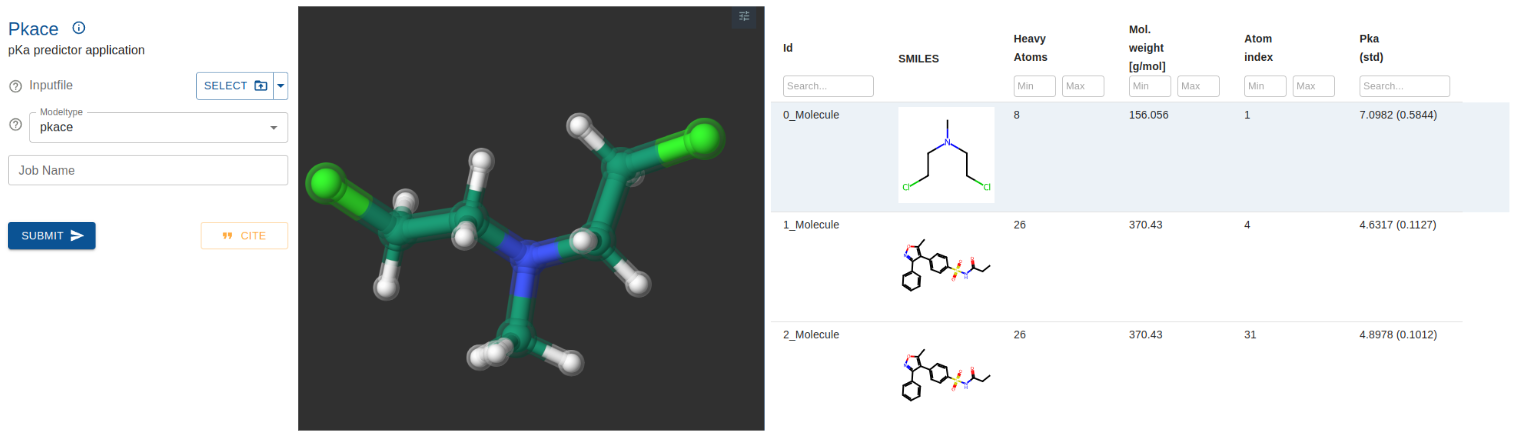}
     \caption{Image of the graphical user interface of the designed application. The graphical interface consists of three main parts: the left-hand side where the inputs to the application are located, the middle screen where the input molecules can be visualized and the right-hand side where the prediction results are shown after completion in a tabular format. The results table consists of the molecule ID as provided in the input file, a 2D representation of the molecule, the number of heavy atoms and the molecular weight of each molecule and the atom index of the found protonation center with the predicted pKa value. In brackets is the standard deviation of the prediction over multiple model replicas.}
     \label{fig:app_gui}
\end{figure*}

In this work, we present a web-based application for pKa predictions of small molecules. We show the graphical user interface to access the application and describe in detail the type of inputs the application supports and the output generated.
We also present the underlying ML model which uses an adapted version of the TensorNet architecture \cite{tensornet} which showed state-of-the-art performance in the prediction of quantum mechanical (QM) molecular properties. The network is both rotationally and translationally equivariant avoiding the need for additional data augmentation. It is built using the TorchMD-net framework \cite{torchmdnet}, a Python library that supports models for applications in neural network potentials and molecular dynamics \cite{torchmdnet}. Different from the existing graph-based networks for molecular property predictions, this network is an adapted version of energy-predicting networks employed for neural network potentials \cite{tensornet,torchmdnet,mace} and works with 3D structures as opposed to other graph-based models that operate on 2D representations. Therefore, the network accepts only atomic elements and atom positions to embed the molecular information around the provided protonation site. This gives the advantage that it can be used for any type of molecular entity and both single as well as multiprotic compounds without any modification to its infrastructure.
The network was trained on datasets of various pKa data compositions to show the influence of different training data distributions and we benchmark the model against publicly available benchmark test sets \cite{acid_base_test,molgpka,novartis_test,literature_test1,literature_test2,literature_test3,literature_test4,literature_test5} and compare its performance against state-of-the-art reference models \cite{ml_meets_pka,molgpka}.

\section{Materials and Methods}
\subsection{Application}

Figure \ref{fig:app_gui} shows the main interface of the application which offers a user-friendly web interface that can be accessed from \hyperlink{https://open.playmolecule.org}{open.playmolecule.org}. It consists of an input panel where the user can submit either single query molecules or entire datasets either in SDF or SMILES format. We offer a selection from one of the pretrained ML models to perform the predictions. The visualization screen in the middle is where the molecules are displayed after loading and on the right-hand side an expanding panel appears after the calculations have been finished where the user can check the generated results.

In the present work, the predictive model is designed to predict micro-pKa values of specific protonation sites within molecules of interest. As molecules can have multiple protonation sites it is necessary to specify these sites for the model so that a prediction can be generated for each. In the current design of the application, while any atom can be accepted as a protonation site for the model, in practice, not all atoms are valid or common sites. To find and select the correct protonation centers, we use a list of common protonation sites SMARTS patterns to detect them automatically.

The predictions are displayed in a sliding window in the main part of the screen of the application as pKa values in the designated column. Each row in the generated results table refers to each specific protonation center in the provided molecules. The location of these centers is given by the atom IDs of the central atom of each identified or provided protonation center. Together with the predicted pKa values, the standard deviation across multiple trained model replicas is provided in brackets which can serve as an approximation of the uncertainty in the generated prediction. The generated results can be further downloaded as an SDF file. In the SDF properties, information about the central atom of each protonation center is stored in the form of an atom ID alongside the predicted pKa value.

\subsection{Model}\label{sec:model}

The model is a graph-based neural network that consists of both rotationally and translationally equivariant layers built using the TorchMD-net framework \cite{torchmdnet}. It is an adapted version of a recently developed, state-of-the-art network named TensorNet \cite{tensornet}. The original network is $O(3)$ equivariant, meaning that it generates the same embeddings when the input molecules are rotated, translated, or mirrored.

\begin{figure}[H]
    \includegraphics[width=1.0\columnwidth]{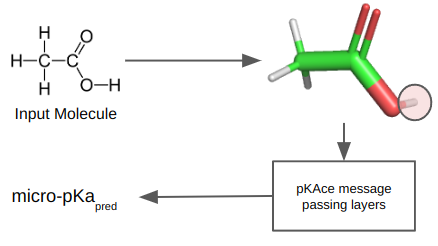}
    \captionsetup{width=1.0\columnwidth}
    \caption{Schematic representation of how input molecules are processed by the application. A 3D conformation of the molecule is first generated followed by the identification of all possible protonation centers. The 3D structure is then passed through the model's message passing layers and pKa predictions for each identified protonation center are generated.}
    \label{fig:schema_app}
\end{figure}

While the $O(3)$ equivariance is important for energy predictions, for molecular properties it is necessary that the network be sensitive to chirality as molecular properties can be different for two stereoisomers. Therefore the original network was adjusted to be $SO(3)$ equivariant to produce different embeddings for chiral molecules. A second modification was done to the final output layers. Instead of taking the sum over all the atoms in the molecule, we modified the network to reduce the atomic embedding to a scalar value of only the central atom of each identified protonation site for which we are predicting the pKa value (see figure \ref{fig:schema_app}). The model was then trained to match the produced scalar output to the experimental pKa of that particular protonation site.

Hyperparameters of the model were optimized through grid search experiments (see SI). To reduce the computational load of running many optimization experiments, selected hyperparameters were modified sequentially, each time updating the hyperparameter values for the tested hyperparameters. To validate the runs a random 80:20 train:validation set split was performed. The batch size was set to the memory limit on one 1080 TI GPU with 12GB of memory. The optimal learning rate was chosen on the basis of the results from the LearningRateFinder function in the Pytorch Lightning package. A complete list of chosen hyperparameters can be found in SI.

Different training data compositions were also studied. As training data can be grouped into experimentally labeled data and data with predicted pKa values by other ML or non-ML methods, it is possible to test how the model trains when using different types of data. For this, we compared a model trained on only experimental values with a model trained on both experimental and predicted values (see SI).

\subsection{Data}\label{sec:data}
For training, both experimental as well as predicted pKa data from Chembl31 \cite{chembl} were tested. Predicted data can be noisier than experimental data due to noise from both the predictive model and the experimental methods used to train it. However, this predicted data constitutes the majority of pKa data found in Chembl31. Therefore, it may still be beneficial to include this type of data in the training process to increase the heterogeneity of the training molecules.

Four established publicly available datasets were taken as external test sets. The first "literature" test set provides a compilation of experimental pKa data from compounds taken from various literature sources \cite{literature_test1,literature_test2,literature_test3,literature_test4,literature_test5}. The "Novartis" test set contains experimental pKa data for some in-house compounds from Novartis \cite{novartis_test}. The "oxy-acids-n-bases" dataset provides experimental pKa data for a range of common acids and bases found in literature \cite{acid_base_test}.  Finally, the "transformation" dataset contains pairs of structurally similar compounds and their experimental pKa data from collected literature sources \cite{molgpka}. This way, it is possible to study the behavior of the trained model on structurally similar compounds.

Consistent with data preparation routines for existing pKa prediction models we follow a similar preparation routine that consists of several steps. First, the molecules are neutralized which means that the molecule is protonated or deprotonated to remove any charges. This is done using the RDKit package by modifying the number of hydrogens and formal charges of the atom that is being neutralized. Sometimes the charge may be locked which means that no protonation or deprotonation is possible. In that case, it will be left unchanged. Further, salts are removed from the molecular structures using the SaltRemover function in RDKit and molecules containing multiple separate structures are also discarded. The molecules are further converted into their canonical tautomers using RDKit and duplicate structures are merged followed by the assignment of the protonation sites. This can be predefined by the user or automatically detected through pattern matching from a SMARTS list of common protonation sites. Both the neutralization and canonicalization of the tautomers are needed due to the SMARTS list of protonation sites being composed starting from such structures which would affect the automatic detection of protonation sites. As we start with the neutral forms of the molecules we organize all the pKa data into acids and bases groups, where acidic protonation sites are sites in their protonated form that contain the hydrogen atom in their neutral form, while basic protonation sites are deprotonated sites that lack this hydrogen in their neutral form. Lastly, the molecules are converted into their 3D forms using the EmbedMolecule function from the RDKit package (see figure \ref{fig:schema_app}).

\section{Results}
\subsection{Training Data Influence}

First, we tested the influence of the training data composition on the model's performance as described in section \ref{sec:model} (see SI). We observed that training on both experimental and predicted pKa data, degrades the performance of the model. This is in line with expectations as probably the predicted pKa data in Chembl31 is too noisy for training. Also, as the MolGpKa model \cite{molgpka} was employed during data preparation (see SI), with the increase of molecular diversity, the risk increases that MolGpka will not be able to accurately predict pKa values for protonation sites and molecules that are too much outside of the data distribution that was used to train the model. Therefore, a higher amount of misassigned protonation centers might be present in this prepared training data which can further contribute to a degradation of our model's performance.

When training on experimental data only, performance improves substantially. We therefore kept the experimental data only as the main training set for the pKAce model.

\subsection{Benchmark Tests}

\begin{figure}[H]
    \centering
    \includegraphics[width=\columnwidth]{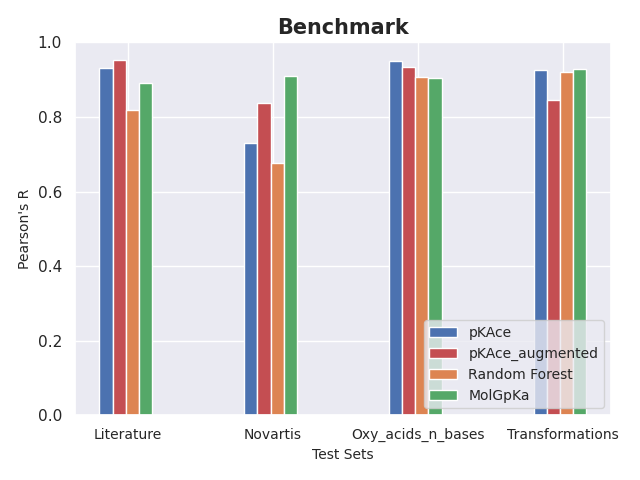}
    \captionsetup{width=1.0\columnwidth}
    \caption{Benchmark results: pKAce models trained on experimental data only with and without data augmentation and tested against established public test sets and compared versus two established predictive models.}
    \label{fig:results_benchmark}
\end{figure}

We further tested the model that was trained on the experimental data only against other ML models. We took both a random forest model \cite{ml_meets_pka} and MolGpka \cite{molgpka}, which is a graph-based neural network model. From the benchmark results (figure \ref{fig:results_benchmark}) we can see that our model reaches a comparable performance to the reference pKa prediction models. Compared to the MolGpka model, this is achieved using only a fraction of the data with which the MolGpka model was trained as its training data consisted of both predicted and experimental pKa data from Chembl. We can also observe a lower performance on the Novartis test set compared to MolGpka. This performance can most probably be attributed to the fact that this test set contains more variability of molecules with different scaffolds, functional groups and protonation site types.

We also observe a lower failure rate of our model compared to the reference models. Hereby our model is able to generate reliable predictions for all of the compounds in the test sets while both reference models fail to process some of the compounds in one or more of the test sets (table \ref{tab:failure_rates}). Therefore, benchmark results were computed only over the successfully processed ones which overestimates the true performance of these models on the test sets, especially for MolGpKa on the Novartis set, as this narrows the test sets down to their respective training data distributions. This means that true performance is lower for both reference models on the indicated test sets and means that our model has more flexibility and adaptability to diverse molecular structures while the reference models are more restricted to specific molecules, scaffolds or functional groups.

\begin{table}[H]
\resizebox{1.0\columnwidth}{!}{%
\begin{tabular}{l|lll}
                                                                     & pKAce & Random Forest & MolGpKa \\ \hline
Literature Test Set                                                  & 0     & 0             & 24      \\
Novartis Test Set                                                    & 0     & 0             & 58      \\
\begin{tabular}[c]{@{}l@{}}Oxy-acids-n-bases\\ Test Set\end{tabular} & 0     & 102           & 0       \\
\begin{tabular}[c]{@{}l@{}}Transformations\\ Test Set\end{tabular}   & 0     & 77            & 0      
\end{tabular}%
}
\captionsetup{width=1.0\columnwidth}
\caption{Number of failed molecules from each benchmark test set during inference by our model and the reference models.}
\label{tab:failure_rates}
\end{table}

\subsection{Data Augmentation Experiments}

Due to the limited amount of training data available, we further investigated two data augmentation techniques to amplify the chemical information contained in the data. In the first technique we deprotonated the acidic sites and protonated the basic sites and added their molecular representations to the opposite pKa data groups to enrich the data with charged molecular forms. In the second technique we generated multiple 3D conformers for each molecule in the training dataset in order to make the model robust for conformational changes of the same molecule input. More information and results for each type of data augmentation are provided in the following subsections.

\subsubsection{Acid-Base Groups Switch}

As described above, in this first data augmentation technique we extended our training data by including both protonated and deprotonated forms of each protonation site and state in our original training data. This should help introduce additional information to the model on the specifics of the proton transfer reaction and improve its performance.

In figure \ref{fig:results_benchmark} are shown the results of the benchmark of the pKAce model trained on the augmented training data following this procedure. We can observe that for most of the test sets, performance remains comparable as when trained on the non-augmented data. However, for the more difficult Novartis test set, performance improves substantially, indicating that additional chemical information helps to improve the generalizability of the model to more complex molecules. As the majority of the molecules in the other test sets are fairly simple and monoprotic, additional information does not provide further improvement in performance as the performance is already at a high level.

\subsubsection{Multiple Conformations}

During training, we further also generate multiple 3D conformers for each molecule. This is necessary to ensure stable predictions when different conformations for a molecule are provided. As the pKa value of a specific protonation site or molecule in aqueous media is an average of the different conformations that it can obtain, it is also a way to describe the training data more realistically. To highlight the advantage of training on multiple conformations we show in SI predictions for different conformers of one of the test compounds predicted with both a pKAce model trained on multiple conformers versus a pKAce model trained on single conformers only. We can observe that training with multiple conformers lowers the variability in predictions for the same molecules and protonation sites compared to training on single conformers only. This is because the different molecular conformations are mapped to the same pKa values during training. Further, including multiple conformations also improves the overall model's performance as can be seen from the lower mean absolute error between the model's predictions and the experimental value.

\section{Conclusion}
In this work we presented an application for small molecule pKa prediction. We introduced the graphical user interface of the application and described its operating details by explaining how to provide the necessary inputs to the application, how the obtained results look like and how these can be downloaded in the supporting formats.
We trained the underlying graph-based neural network model with an architecture adapted from energy prediction models and show that our model is able to meet comparable performance to reference models despite being trained on fewer samples. Hereby we show that extra care needs to be taken during data preparation of pKa data coming from computed sources. We also show that our model has a considerably lower failure rate during prediction compared to reference models due to its flexible architectural design and is able to accept a wider range of diverse molecular structures without any modification to its underlying architecture. We leave further exploration of this potential to future work in order to expand the range of molecular entities and pKa prediction cases of our model through training on additional data for these entities. We also show that data augmentations rooted in chemical logic of the protonation/deprotonation event or the inclusion of the different 3D conformational states that a molecule can adopt in aqueous solutions, helps to further improve the performance of the trained models and maximizes the information utilization that is present in the training data. We further make both the pKace and MolGpKa models available in the application.

Because of the flexible architecture and minimal atomic featurization, the model can easily be trained on and used for any type of molecular structure. This is an interesting advantage as currently ML models for pKa predictions are generally divided between models handling small molecules or larger molecular structures such as peptides or proteins. Therefore, merging both types of data would not only enable the development of a universal molecular pKa prediction model but could also improve performance further through a larger and more diverse training set of molecular structures. Apart from that, recent work \cite{epik} has also shown improved performance when including the different tautomeric forms of the molecules rather than the most energetically favorable one. Just as when adding both forms of each protonation state, this expands the diversity of the training data. Both are interesting directions for further exploration which we leave for further study.

\section{Supporting Information}
"Hyperparameter selection", "data analysis" and "additional results", additional information for the sections; Tables S1-S2 information on tested and chosen model hyperparameters; Tables S3-S4 general statistics of the train and test compounds; Figures S1-S5 additional statistics of the train compounds; Figures S6-S24 additional statistics of the test compounds; Figure S25 mean absolute error performances on the benchmark test sets; Table S5 results of the conformers test (PDF)

\section{Acknowledgements} 
This work was supported by the Industrial Doctorates Plan of the Secretariat of Universities and Research of the Department of Economy and Knowledge of the Generalitat of Catalonia.
This project has received funding from the Torres-Quevedo Programme from the Spanish National Agency for Research (PTQ2021-011670 / AEI / 10.13039/501100011033).

\section{Data and Software Availability}
pKAce is available free of charge at \hyperlink{https://open.playmolecule.org}{open.playmolecule.com}. The databases used for training (CHEMBL31 \cite{chembl}) and validating (Literature \cite{literature_test1,literature_test2,literature_test3,literature_test4,literature_test5}, Novartis \cite{novartis_test}, Oxy-acids-n-bases \cite{acid_base_test}, Transformations \cite{molgpka}) the models are publicly available. Further we provide the prepared test sets molecules ready to download as SDF files in the examples section of the application GUI and we make the different pKAce models tested in this work available in the application. We provide an easy and straightforward way to retest the test sets compounds with the different pKAce models using the public application GUI. The data to produce the GUI images is also included in the examples section of the application.

\printbibliography

@misc{tensornet,
      title={TensorNet: Cartesian Tensor Representations for Efficient Learning of Molecular Potentials}, 
      author={Guillem Simeon and Gianni de Fabritiis},
      year={2023},
      eprint={2306.06482},
      archivePrefix={arXiv},
      primaryClass={cs.LG}
}

@article{kdeep,
    author = {Jiménez, José and Škalič, Miha and Martínez-Rosell, Gerard and De Fabritiis, Gianni},
    title = {KDEEP: Protein–Ligand Absolute Binding Affinity Prediction via 3D-Convolutional Neural Networks},
    journal = {Journal of Chemical Information and Modeling},
    volume = {58},
    number = {2},
    pages = {287-296},
    year = {2018},
    doi = {10.1021/acs.jcim.7b00650},
    note ={PMID: 29309725},
    URL = {https://doi.org/10.1021/acs.jcim.7b00650},
    eprint = {https://doi.org/10.1021/acs.jcim.7b00650}
}

@article{cardio_tox,
    author = {Chen, Yuanting and Yu, Xinxin and Li, Weihua and Tang, Yun and Liu, Guixia},
    title = {In silico prediction of hERG blockers using machine learning and deep learning approaches},
    journal = {Journal of Applied Toxicology},
    volume = {n/a},
    number = {n/a},
    pages = {},
    doi = {https://doi.org/10.1002/jat.4477},
    url = {https://analyticalsciencejournals.onlinelibrary.wiley.com/doi/abs/10.1002/jat.4477},
    eprint = {https://analyticalsciencejournals.onlinelibrary.wiley.com/doi/pdf/10.1002/jat.4477}
}

@article{torchmdnet,
  author = {Philipp Th{\"{o}}lke and Gianni De Fabritiis},
  title = {TorchMD-NET: Equivariant Transformers for Neural Network based Molecular Potentials},
  journal = {CoRR},
  volume = {abs/2202.02541},
  year = {2022},
  url = {https://arxiv.org/abs/2202.02541},
  eprinttype = {arXiv},
  eprint = {2202.02541},
  bibsource = {dblp computer science bibliography, https://dblp.org}
}

@article{chembl,
    author = {Mendez, David and Gaulton, Anna and Bento, A Patrícia and Chambers, Jon and De Veij, Marleen and Félix, Eloy and Magariños, María Paula and Mosquera, Juan F and Mutowo, Prudence and Nowotka, Michał and Gordillo-Marañón, María and Hunter, Fiona and Junco, Laura and Mugumbate, Grace and Rodriguez-Lopez, Milagros and Atkinson, Francis and Bosc, Nicolas and Radoux, Chris J and Segura-Cabrera, Aldo and Hersey, Anne and Leach, Andrew R},
    title = "{ChEMBL: towards direct deposition of bioassay data}",
    journal = {Nucleic Acids Res.},
    volume = {47},
    number = {D1},
    pages = {D930-D940},
    year = {2018},
    month = {11},
    issn = {0305-1048},
    doi = {10.1093/nar/gky1075},
    url = {https://doi.org/10.1093/nar/gky1075},
    eprint = {https://academic.oup.com/nar/article-pdf/47/D1/D930/27437436/gky1075.pdf},
}

@article{epik,
author = {Johnston, Ryne C. and Yao, Kun and Kaplan, Zachary and Chelliah, Monica and Leswing, Karl and Seekins, Sean and Watts, Shawn and Calkins, David and Chief Elk, Jackson and Jerome, Steven V. and Repasky, Matthew P. and Shelley, John C.},
title = {Epik: pKa and Protonation State Prediction through Machine Learning},
journal = {J. of Chem. Theory and Comp.},
volume = {19},
number = {8},
pages = {2380-2388},
year = {2023},
doi = {10.1021/acs.jctc.3c00044},
note ={PMID: 37023332},
URL = {https://doi.org/10.1021/acs.jctc.3c00044},
eprint = {https://doi.org/10.1021/acs.jctc.3c00044}
}

@article{prot_state_importance, title={The role of protonation states in ligand-receptor recognition and binding}, volume={19}, DOI={10.2174/1381612811319230004}, number={23}, journal={Curr. Pharm. Des.}, author={Petukh, Marharyta and Stefl, Shannon and Alexov, Emil}, year={2013}, pages={4182–4190}}

@article{bindscope,
    author = {Skalic, Miha and Martínez-Rosell, Gerard and Jiménez, José and De Fabritiis, Gianni},
    title = "{PlayMolecule BindScope: large scale CNN-based virtual screening on the web}",
    journal = {Bioinf.},
    volume = {35},
    number = {7},
    pages = {1237-1238},
    year = {2018},
    month = {08},
    issn = {1367-4803},
    doi = {10.1093/bioinformatics/bty758},
    url = {https://doi.org/10.1093/bioinformatics/bty758},
    eprint = {https://academic.oup.com/bioinformatics/article-pdf/35/7/1237/48968405/bioinformatics\_35\_7\_1237.pdf},
}

@article{GGL-Score,
title = {Geometric graph learning with extended atom-types features for protein-ligand binding affinity prediction},
journal = {Computers in Biology and Medicine},
volume = {164},
pages = {107250},
year = {2023},
issn = {0010-4825},
doi = {https://doi.org/10.1016/j.compbiomed.2023.107250},
url = {https://www.sciencedirect.com/science/article/pii/S0010482523007151},
author = {Md Masud Rana and Duc Duy Nguyen}
}

@misc{ml_dd_overview,
      title={Machine Learning Small Molecule Properties in Drug Discovery}, 
      author={Nikolai Schapin and Maciej Majewski and Alejandro Varela and Carlos Arroniz and Gianni De Fabritiis},
      year={2023},
      eprint={2308.12354},
      archivePrefix={arXiv},
      primaryClass={q-bio.BM}
}

@article{sol_ml,
author = {Broccatelli, Fabio and Trager, Richard and Reutlinger, Michael and Karypis, George and Li, Mufei},
title = {Benchmarking Accuracy and Generalizability of Four Graph Neural Networks Using Large In Vitro ADME Datasets from Different Chemical Spaces},
journal = {Mol. Inf.},
volume = {41},
number = {8},
pages = {2100321},
doi = {https://doi.org/10.1002/minf.202100321},
url = {https://onlinelibrary.wiley.com/doi/abs/10.1002/minf.202100321},
eprint = {https://onlinelibrary.wiley.com/doi/pdf/10.1002/minf.202100321},
year = {2022}
}

@article{lipo_ml,
author = {Win, Zaw-Myo and Cheong, Allen M. Y. and Hopkins, W. Scott},
title = {Using Machine Learning To Predict Partition Coefficient (Log P) and Distribution Coefficient (Log D) with Molecular Descriptors and Liquid Chromatography Retention Time},
journal = {J. of Chem. Inf. and Model.},
volume = {63},
number = {7},
pages = {1906-1913},
year = {2023},
doi = {10.1021/acs.jcim.2c01373},
note ={PMID: 36926888},
URL = {https://doi.org/10.1021/acs.jcim.2c01373},
eprint = {https://doi.org/10.1021/acs.jcim.2c01373}
}

@article{ggl_tox,
author = {Jiang, Jian and Wang, Rui and Wei, Guo-Wei},
title = {GGL-Tox: Geometric Graph Learning for Toxicity Prediction},
journal = {Journal of Chemical Information and Modeling},
volume = {61},
number = {4},
pages = {1691-1700},
year = {2021},
doi = {10.1021/acs.jcim.0c01294},
note ={PMID: 33719422},
URL = {https://doi.org/10.1021/acs.jcim.0c01294},
eprint = {https://doi.org/10.1021/acs.jcim.0c01294}
}

@Article{adme1,
AUTHOR = {Li, Xinkang and Tang, Lijun and Li, Zeying and Qiu, Dian and Yang, Zhuoling and Li, Baoqiong},
TITLE = {Prediction of ADMET Properties of Anti-Breast Cancer Compounds Using Three Machine Learning Algorithms},
JOURNAL = {Mol.},
VOLUME = {28},
YEAR = {2023},
NUMBER = {5},
ARTICLE-NUMBER = {2326},
URL = {https://www.mdpi.com/1420-3049/28/5/2326},
PubMedID = {36903569},
ISSN = {1420-3049},
DOI = {10.3390/molecules28052326}
}

@article{admet2,
author = {Zhou, Yadi and Cahya, Suntara and Combs, Steven A. and Nicolaou, Christos A. and Wang, Jibo and Desai, Prashant V. and Shen, Jie},
title = {Exploring Tunable Hyperparameters for Deep Neural Networks with Industrial ADME Data Sets},
journal = {J. of Chem. Inf. and Model.},
volume = {59},
number = {3},
pages = {1005-1016},
year = {2019},
doi = {10.1021/acs.jcim.8b00671},
URL = {https://doi.org/10.1021/acs.jcim.8b00671},
eprint = {https://doi.org/10.1021/acs.jcim.8b00671}
}

@article{reinvent,
author = {Blaschke, Thomas and Arús-Pous, Josep and Chen, Hongming and Margreitter, Christian and Tyrchan, Christian and Engkvist, Ola and Papadopoulos, Kostas and Patronov, Atanas},
title = {REINVENT 2.0: An AI Tool for De Novo Drug Design},
journal = {Journal of Chemical Information and Modeling},
volume = {60},
number = {12},
pages = {5918-5922},
year = {2020},
doi = {10.1021/acs.jcim.0c00915},
note ={PMID: 33118816},
URL = {https://doi.org/10.1021/acs.jcim.0c00915},
eprint = {https://doi.org/10.1021/acs.jcim.0c00915}
}

@article{ml_meets_pka,
  doi = {10.12688/f1000research.22090.2},
  url = {https://doi.org/10.12688/f1000research.22090.2},
  year = {2020},
  month = apr,
  publisher = {F1000 Research Ltd},
  volume = {9},
  pages = {113},
  author = {Marcel Baltruschat and Paul Czodrowski},
  title = {Machine learning meets {pKa}},
  journal = {F1000Research}
}

@article{pka_rooted_top_fp,
author = {Lu, Yipin and Anand, Shankara and Shirley, William and Gedeck, Peter and Kelley, Brian P. and Skolnik, Suzanne and Rodde, Stephane and Nguyen, Mai and Lindvall, Mika and Jia, Weiping},
title = {Prediction of pKa Using Machine Learning Methods with Rooted Topological Torsion Fingerprints: Application to Aliphatic Amines},
journal = {J. of Chem. Inf. and Model.},
volume = {59},
number = {11},
pages = {4706-4719},
year = {2019},
doi = {10.1021/acs.jcim.9b00498},
note ={PMID: 31647238},
URL = {https://doi.org/10.1021/acs.jcim.9b00498},
eprint = {https://doi.org/10.1021/acs.jcim.9b00498}
}

@article{multi-instance_pka,
    author = {Xiong, Jiacheng and Li, Zhaojun and Wang, Guangchao and Fu, Zunyun and Zhong, Feisheng and Xu, Tingyang and Liu, Xiaomeng and Huang, Ziming and Liu, Xiaohong and Chen, Kaixian and Jiang, Hualiang and Zheng, Mingyue},
    title = "{Multi-instance learning of graph neural networks for aqueous pKa prediction}",
    journal = {Bioinform.},
    volume = {38},
    number = {3},
    pages = {792-798},
    year = {2021},
    month = {10},
    doi = {10.1093/bioinformatics/btab714},
    url = {https://doi.org/10.1093/bioinformatics/btab714},
    eprint = {https://academic.oup.com/bioinformatics/article-pdf/38/3/792/49007282/btab714.pdf},
}

@article{molgpka,
author = {Pan, Xiaolin and Wang, Hao and Li, Cuiyu and Zhang, John Z. H. and Ji, Changge},
title = {MolGpka: A Web Server for Small Molecule pKa Prediction Using a Graph-Convolutional Neural Network},
journal = {J. of Chem. Inf. and Model.},
volume = {61},
number = {7},
pages = {3159-3165},
year = {2021},
doi = {10.1021/acs.jcim.1c00075},
note ={PMID: 34251213},
URL = {https://doi.org/10.1021/acs.jcim.1c00075},
eprint = {https://doi.org/10.1021/acs.jcim.1c00075}
}

@article{challenges_pka,
title = {Machine learning methods for pKa prediction of small molecules: Advances and challenges},
journal = {Drug Discovery Today},
volume = {27},
number = {12},
pages = {103372},
year = {2022},
issn = {1359-6446},
doi = {https://doi.org/10.1016/j.drudis.2022.103372},
url = {https://www.sciencedirect.com/science/article/pii/S1359644622003658},
author = {Jialu Wu and Yu Kang and Peichen Pan and Tingjun Hou}
}

@article{chemaxon,
author = {Lee, Adam C. and Crippen, Gordon M.},
title = {Predicting pKa},
journal = {Journal of Chemical Information and Modeling},
volume = {49},
number = {9},
pages = {2013-2033},
year = {2009},
doi = {10.1021/ci900209w},
note ={PMID: 19702243},
URL = {https://doi.org/10.1021/ci900209w},
eprint = {https://doi.org/10.1021/ci900209w}
}

@inproceedings{mace,
title={{MACE}: Higher Order Equivariant Message Passing Neural Networks for Fast and Accurate Force Fields},
author={Ilyes Batatia and David Peter Kovacs and Gregor N. C. Simm and Christoph Ortner and Gabor Csanyi},
booktitle={Advances in Neural Information Processing Systems},
editor={Alice H. Oh and Alekh Agarwal and Danielle Belgrave and Kyunghyun Cho},
year={2022},
url={https://openreview.net/forum?id=YPpSngE-ZU}
}

@article{acid_base_test,
author = {Yu, Haiying and Kühne, Ralph and Ebert, Ralf-Uwe and Schüürmann, Gerrit},
title = {Comparative Analysis of QSAR Models for Predicting pKa of Organic Oxygen Acids and Nitrogen Bases from Molecular Structure},
journal = {Journal of Chemical Information and Modeling},
volume = {50},
number = {11},
pages = {1949-1960},
year = {2010},
doi = {10.1021/ci100306k},
note ={PMID: 21033677},
URL = {https://doi.org/10.1021/ci100306k},
eprint = {https://doi.org/10.1021/ci100306k}
}

@article{novartis_test,
author = {Liao, Chenzhong and Nicklaus, Marc C.},
title = {Comparison of Nine Programs Predicting pKa Values of Pharmaceutical Substances},
journal = {Journal of Chemical Information and Modeling},
volume = {49},
number = {12},
pages = {2801-2812},
year = {2009},
doi = {10.1021/ci900289x},
note ={PMID: 19961204},
URL = {https://doi.org/10.1021/ci900289x},
eprint = {https://doi.org/10.1021/ci900289x}
}

@inbook{literature_test1,
publisher = {John Wiley \& Sons, Ltd},
isbn = {9781118286067},
title = {pKa Determination},
booktitle = {Absorption and Drug Development},
chapter = {3},
pages = {31-173},
doi = {https://doi.org/10.1002/9781118286067.ch3},
url = {https://onlinelibrary.wiley.com/doi/abs/10.1002/9781118286067.ch3},
eprint = {https://onlinelibrary.wiley.com/doi/pdf/10.1002/9781118286067.ch3},
year = {2012}
}

@article{literature_test2,
author = {Morgenthaler, Martin and Schweizer, Eliane and Hoffmann-Röder, Anja and Benini, Fausta and Martin, Rainer E. and Jaeschke, Georg and Wagner, Björn and Fischer, Holger and Bendels, Stefanie and Zimmerli, Daniel and Schneider, Josef and Diederich, François and Kansy, Manfred and Müller, Klaus},
title = {Predicting and Tuning Physicochemical Properties in Lead Optimization: Amine Basicities},
journal = {ChemMedChem},
volume = {2},
number = {8},
pages = {1100-1115},
doi = {https://doi.org/10.1002/cmdc.200700059},
url = {https://chemistry-europe.onlinelibrary.wiley.com/doi/abs/10.1002/cmdc.200700059},
eprint = {https://chemistry-europe.onlinelibrary.wiley.com/doi/pdf/10.1002/cmdc.200700059},
year = {2007}
}

@article{literature_test3,
  doi = {10.1007/s11095-005-6246-8},
  url = {https://doi.org/10.1007/s11095-005-6246-8},
  year = {2005},
  month = aug,
  publisher = {Springer Science and Business Media {LLC}},
  volume = {22},
  number = {9},
  pages = {1454--1460},
  author = {Feng Luan and Weiping Ma and Haixia Zhang and Xiaoyun Zhang and Mancang Liu and Zhide Hu and Botao Fan},
  title = {Prediction of {pKa} for Neutral and Basic Drugs Based on Radial Basis Function Neural Networks and the Heuristic Method},
  journal = {Pharmaceutical Research}
}

@article{literature_test4,
title = {Automated techniques in pKa determination: Low, medium and high-throughput screening methods},
journal = {Drug Discovery Today: Technologies},
volume = {27},
pages = {49-58},
year = {2018},
note = {Physicochemical characterisation in drug discovery},
issn = {1740-6749},
doi = {https://doi.org/10.1016/j.ddtec.2018.04.001},
url = {https://www.sciencedirect.com/science/article/pii/S1740674917300367},
author = {Christophe Dardonville}
}

@article{literature_test5,
  doi = {10.4137/aci.s12304},
  url = {https://doi.org/10.4137/aci.s12304},
  year = {2013},
  month = jan,
  publisher = {{SAGE} Publications},
  volume = {8},
  pages = {ACI.S12304},
  author = {Jetse Reijenga and Arno van Hoof and Antonie van Loon and Bram Teunissen},
  title = {Development of Methods for the Determination of pKa Values},
  journal = {Analytical Chemistry Insights}
}
\end{multicols}
\newpage
\tableofcontents

\end{document}


\maketitle

\tableofcontents
\addcontentsline{toc}{section}{List of Figures}
\listoffigures
\addcontentsline{toc}{section}{List of Tables}
\listoftables\newpage

\section{Hyperparameter Optimization}

\begin{table}[H]
\resizebox{\textwidth}{!}{%
\begin{tabular}{ll}
\textbf{Tested Hyperparameter}         & \textbf{Tested Values}  \\ \hline
Embedding Filter                       & 32, 64, 128, 256        \\
Number of interactions                 & 0, 1, 2                 \\
Upper atomic cutoff                    & 2, 5, 10                \\
Number of distance expansion functions & 32, 64, 128, 256        \\
Distance expansion function type       & Expnorm, Gaussian       \\
Learning Rate Scheduler Type & ReduceLrOnPlateau, ReduceLrOnPlateau with learning rate warmup \\
Learning Rate Warmup Steps             & 500, 1000, 20000, 80000 \\
Learning Rate Scheduler Patience       & 26000, 52000           
\end{tabular}%
}
\caption{Tested hyperparameters and values.}
\label{tab:hyperparam_opt}
\end{table}

\begin{table}[H]
\resizebox{\textwidth}{!}{%
\begin{tabular}{ll}
\textbf{Hyperparameter}                & \textbf{Chosen Value} \\ \hline
Embedding Filter                       & 128                   \\
Number of interactions                 & 1                     \\
Lower atomic cutoff                    & 0                     \\
Upper atomic cutoff                    & 5                     \\
Number of distance expansion functions & 32                    \\
Distance expansion function type       & Expnorm               \\
Learning rate                          & 0.0001                \\
Learning rate scheduler type           & ReduceLrOnPlateau     \\
Learning rate scheduler patience       & 26000 steps           \\
Scheduler factor                       & 0.97                  \\
Weight decay                           & 0.0                   \\
Activation function                    & SiLu                  \\
Batch size                             & 64                   
\end{tabular}%
}
\caption{Chosen hyperparameter values.}
\label{tab:hyperparam_chosen}
\end{table}

\section{Datasets}
\subsection{Data preparation}

Training data was obtained from CHEMBL31 using the same filters as described in \cite{molgpka}. The molecules were further prepared as described in the main text. In order to find and select the correct protonation sites, predictions of the MolGpKa model \cite{molgpka} were used by selecting the protonation site with the predicted pKa closest to the provided pKa value. For some of the data, protonation site estimation results were also available using ChemAxon \cite{chemaxon}. For this data, only data was used were both MolGpKa and ChemAxon agree on the protonation site. In case no additional reference was available, molecules for which MolGpKa predicted multiple protonation sites where more than one protonation site had a predicted value closer than 1 unit difference, were discarded.

\subsection{Training Dataset Statistics}

\begin{table}[H]
\resizebox{\textwidth}{!}{%
\begin{tabular}{lll}
 & \begin{tabular}[c]{@{}l@{}}Experimental pKa\\ Values\end{tabular} \\
Number of Molecules                                                         & 7094  \\
\begin{tabular}[c]{@{}l@{}}Number of Monoprotic\\ Molecules\end{tabular}    & 6625  \\
\begin{tabular}[c]{@{}l@{}}Number of Multiprotic\\ Molecules\end{tabular}   & 469   \\
\begin{tabular}[c]{@{}l@{}}Number of Acid\\ Protonation Sites\end{tabular}  & 2658  \\
\begin{tabular}[c]{@{}l@{}}Number of Basic\\ Protonation Sites\end{tabular} & 4905  \\
\begin{tabular}[c]{@{}l@{}}Number of Unique\\ Scaffolds\end{tabular}        & 3039 
\end{tabular}%
}
\caption{Statistics on the training dataset of experimental pKa values.}
\label{tab:train_data_stats}
\end{table}

\begin{figure}[H]
\centering
     \includegraphics[width=1.0\textwidth]{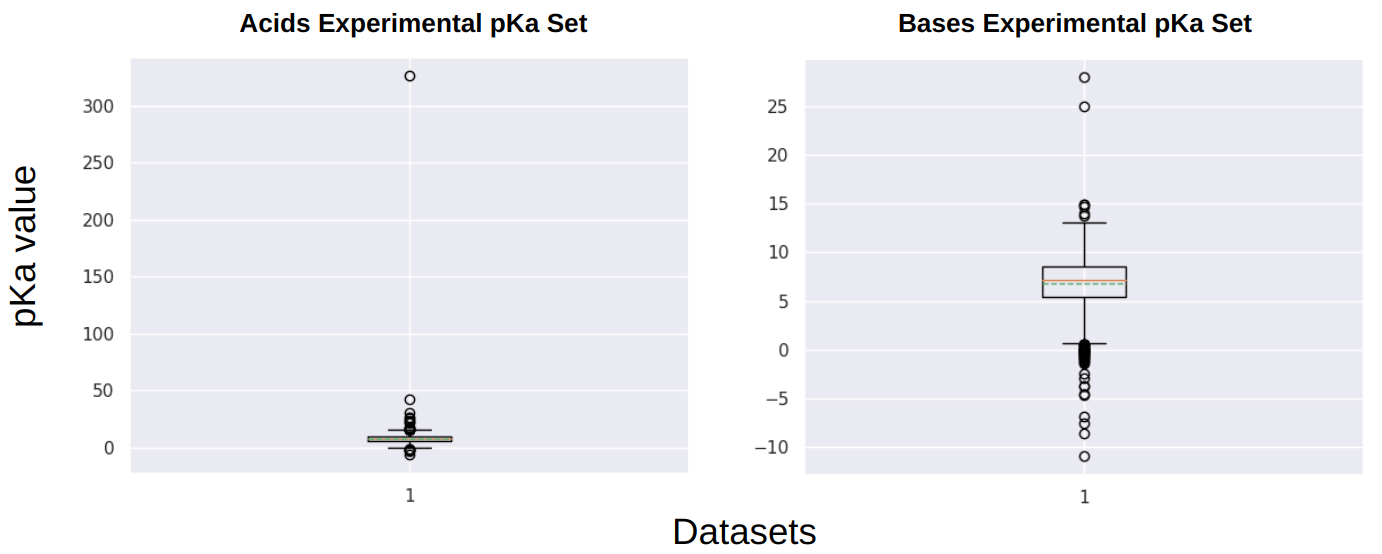}
      \caption{Boxplots for acid and base protonation subsets of the experimental pKas training dataset.}
       \label{fig:boxplot_train}
\end{figure}

\begin{figure}[H]
\centering
     \includegraphics[width=1.0\textwidth]{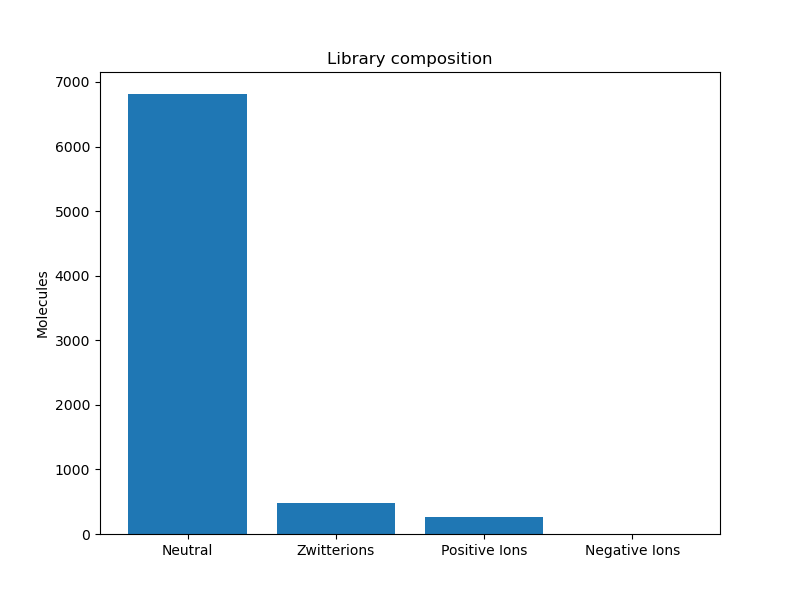}
      \caption{Molecule types in the experimental pKa values train set.}
       \label{fig:exp_pka_train_libcomp}
\end{figure}

\begin{figure}[H]
\centering
     \includegraphics[width=1.0\textwidth]{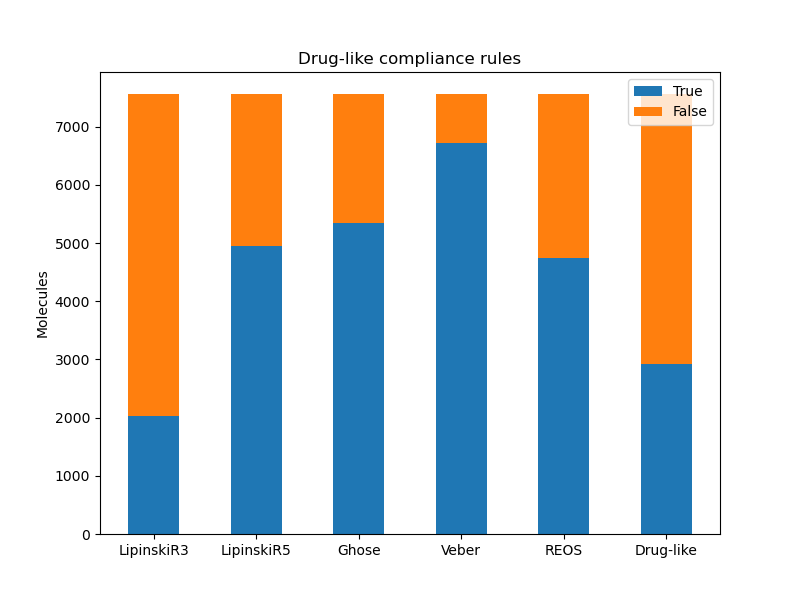}
      \caption{Number of molecules compliant with common drugability rules in the experimental pKa values train set.}
       \label{fig:exp_pka_train_compliance}
\end{figure}

\begin{figure}[H]
\centering
     \includegraphics[width=1.0\textwidth]{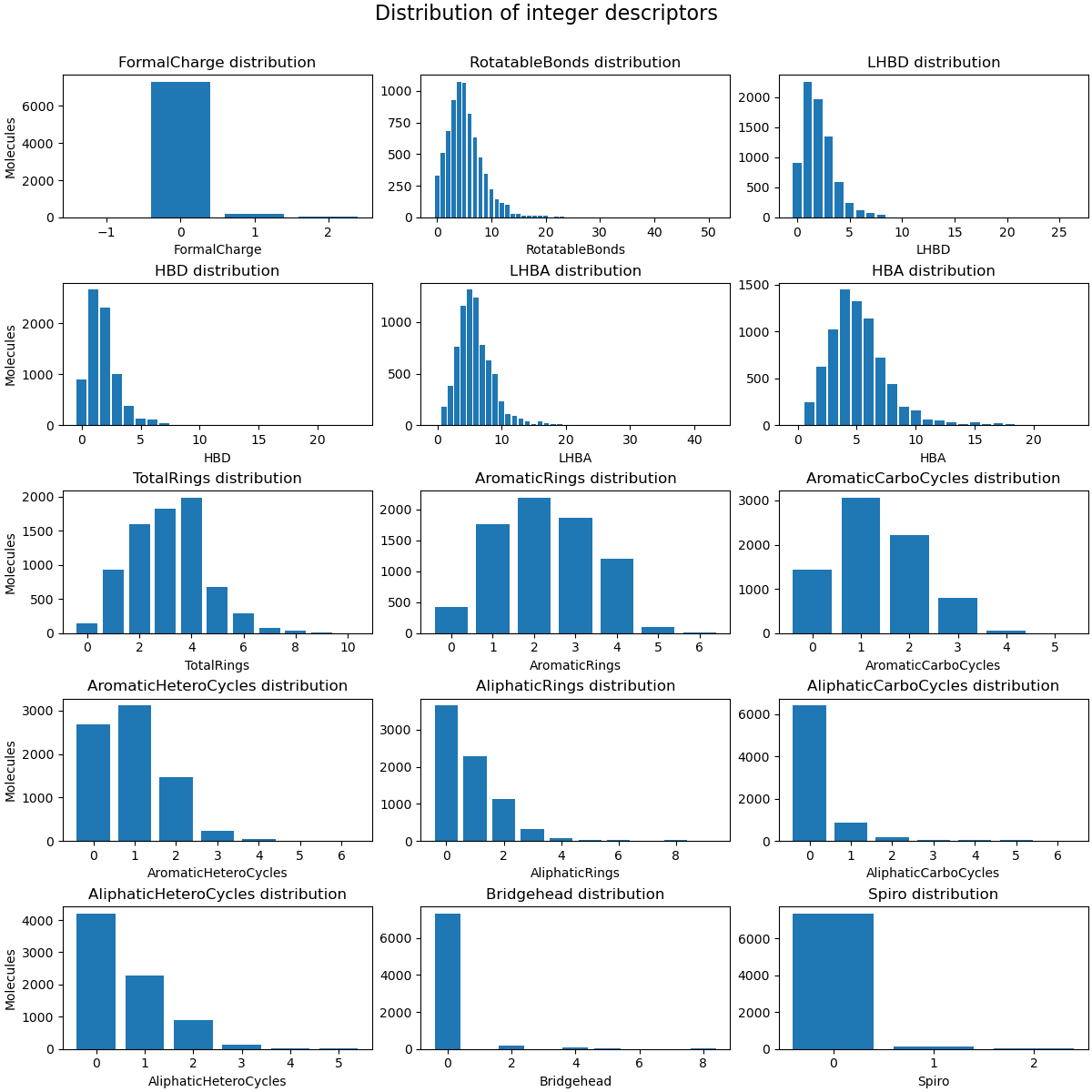}
      \caption{Additional integer dataset descriptors for the experimental pKa values train set. Abbreviations: LHBD=Lipinski Hydrogen Bond Donors, HBD=Hydrogen Bond Donors, LHBA=Lipinski Hydrogen Bond Acceptors, HBA=Hydrogen Bond Acceptors}
       \label{fig:exp_pka_train_intdesc}
\end{figure}

\begin{figure}[H]
\centering
     \includegraphics[width=1.0\textwidth]{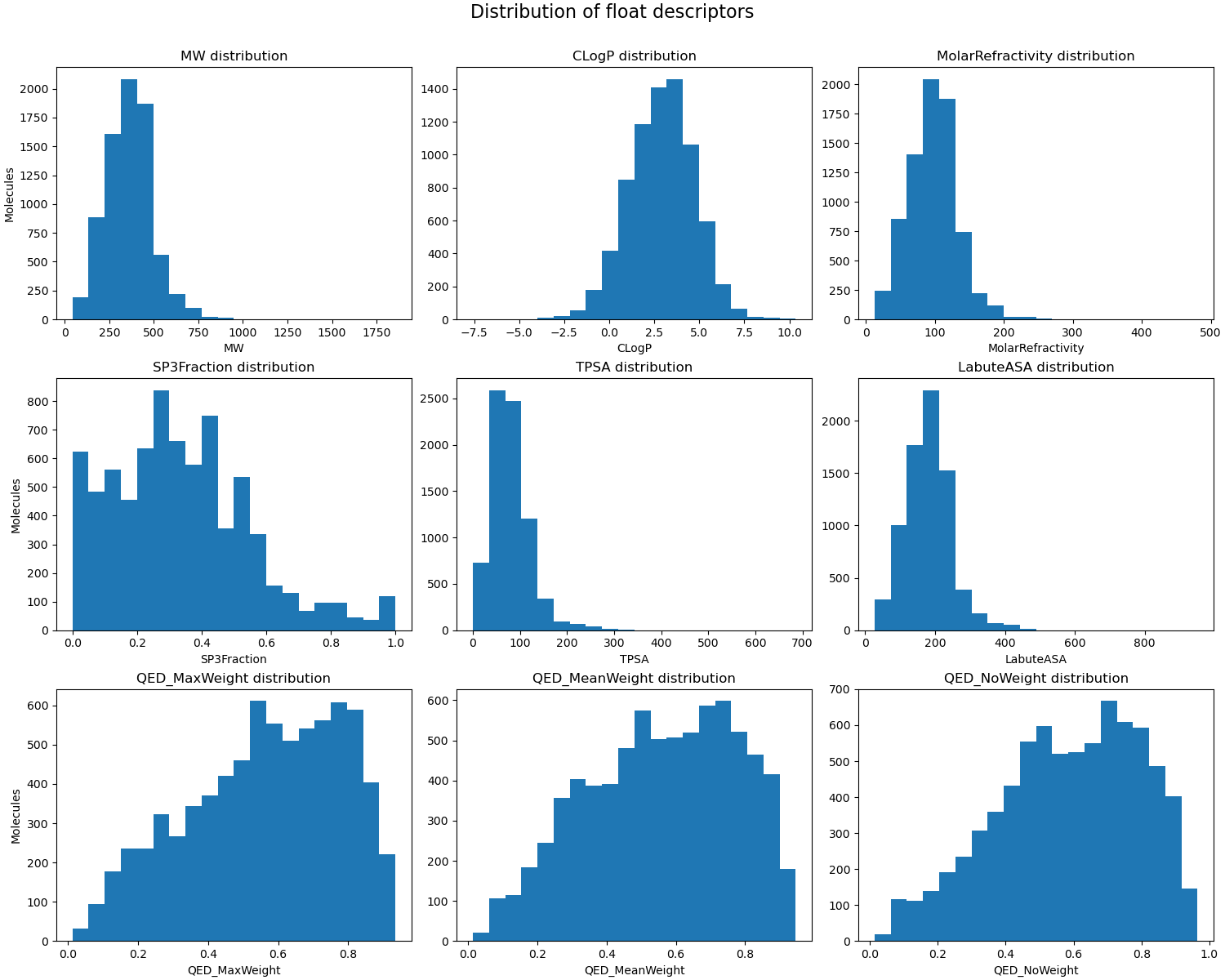}
      \caption{Additional float dataset descriptors for the experimental pKa values train set. Abbreviations: TPSA=Topological Polar Surface Area, LabuteASA=Labute Accessible Surface Area, QED=Quantitative Estimate of Druglikeness}
       \label{fig:exp_pka_train_floatdesc}
\end{figure}

\subsection{Benchmark Test Datasets Statistics}

\begin{table}[H]
\resizebox{\textwidth}{!}{%
\begin{tabular}{lllll}
 &
  \begin{tabular}[c]{@{}l@{}}Literature Test\\ Set\end{tabular} &
  Novartis Test Set &
  \begin{tabular}[c]{@{}l@{}}Oxy-bases-n-acids\\ Test Set\end{tabular} &
  \begin{tabular}[c]{@{}l@{}}Transformations\\ Test Set\end{tabular} \\
Number of Molecules                                                          & 111   & 249  & 858   & 2539  \\
\begin{tabular}[c]{@{}l@{}}Number of Monoprotic\\ Molecules\end{tabular}     & 111   & 245  & 788   & 2443  \\
\begin{tabular}[c]{@{}l@{}}Number of Multiprotic\\ Molecules\end{tabular}    & 0     & 4    & 70    & 96    \\
\begin{tabular}[c]{@{}l@{}}Number of Acid\\ Protonation Sites\end{tabular}   & 0     & 42   & 561   & 1489  \\
\begin{tabular}[c]{@{}l@{}}Number of Basic\\ Protonation Sites\end{tabular}  & 111   & 211  & 367   & 1146  \\
\begin{tabular}[c]{@{}l@{}}Number of Unique\\ Scaffolds\end{tabular}         & 69    & 200  & 237   & 704   \\
\begin{tabular}[c]{@{}l@{}}Percent Scaffolds not\\ in Train Set\end{tabular} & 68.12 & 86.0 & 75.53 & 65.06
\end{tabular}%
}
\caption{Statistics on the benchmark test sets.}
\label{tab:test_data_stats}
\end{table}

\begin{figure}[H]
\centering
     \includegraphics[width=1.0\textwidth]{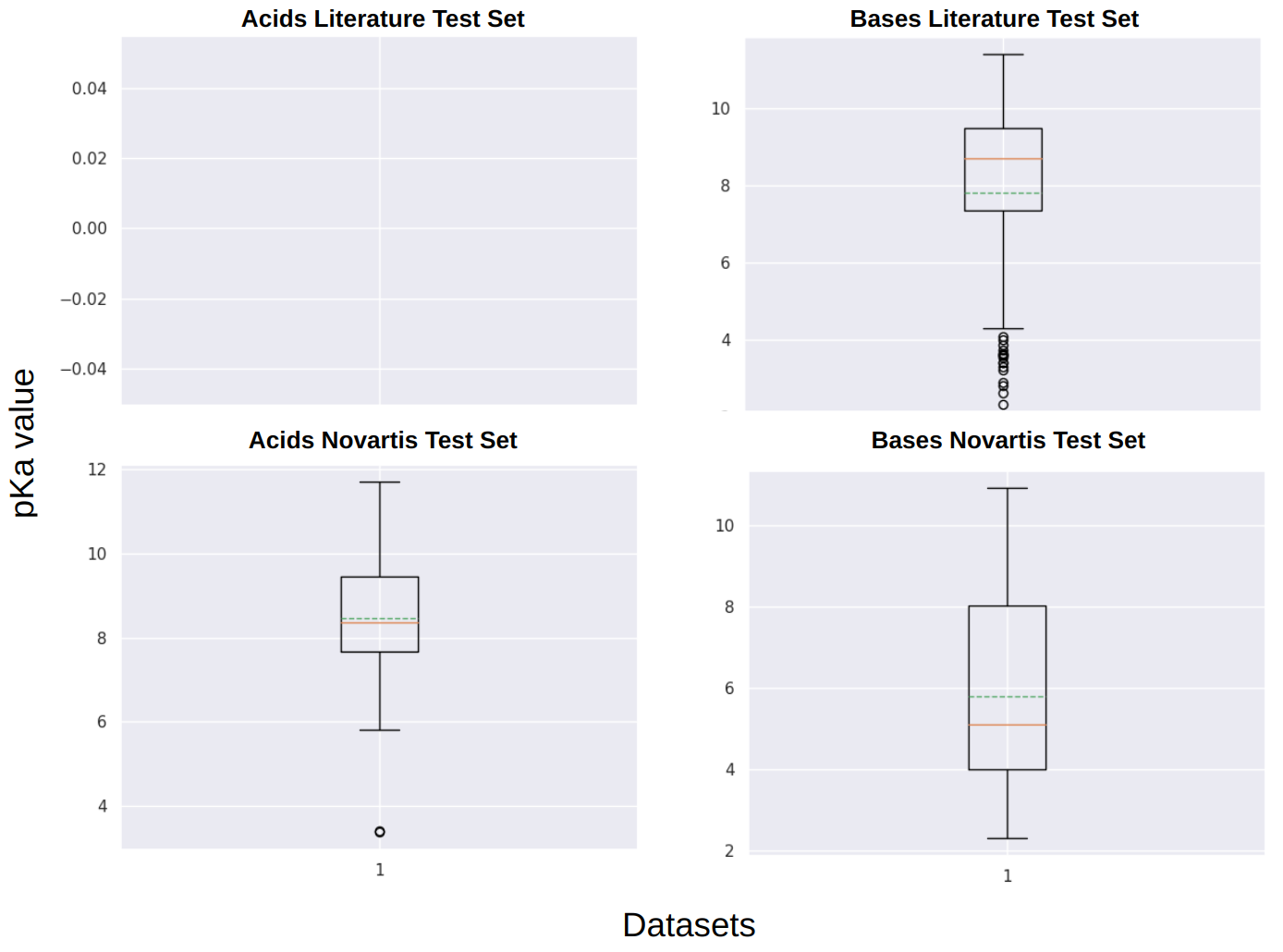}
      \caption{Boxplots for acid and base protonation subsets of the Literature and Novartis test sets.}
       \label{fig:boxplot_lit_nov_test}
\end{figure}

\begin{figure}[H]
\centering
     \includegraphics[width=1.0\textwidth]{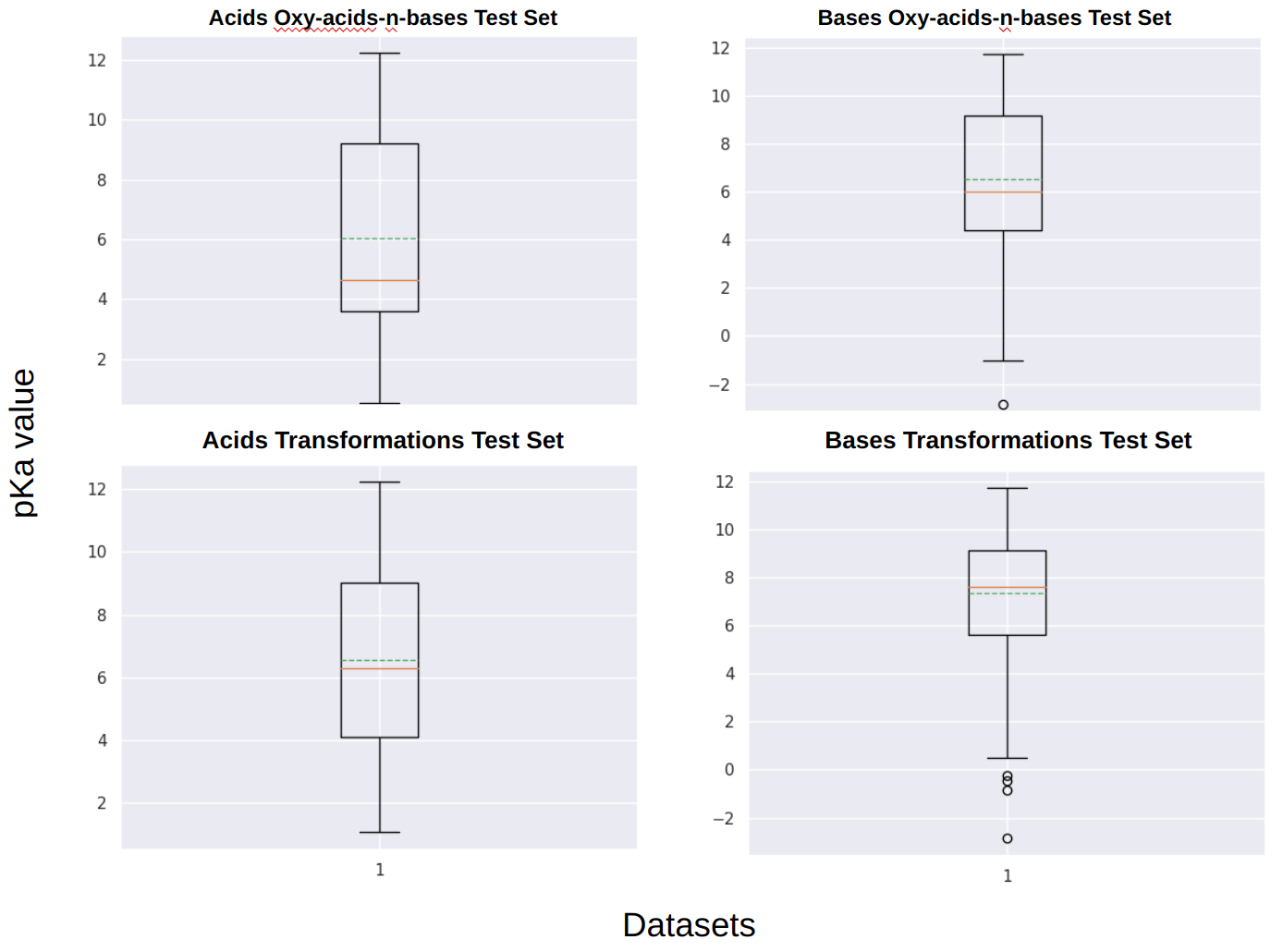}
      \caption{Boxplots for acid and base protonation subsets of the Oxy-acids-n-bases and Transformations test sets.}
       \label{fig:boxplot_oxy_transform_test}
\end{figure}

\subsubsection{Literature Test Set Additional Statistics}

\begin{figure}[H]
\centering
     \includegraphics[width=1.0\textwidth]{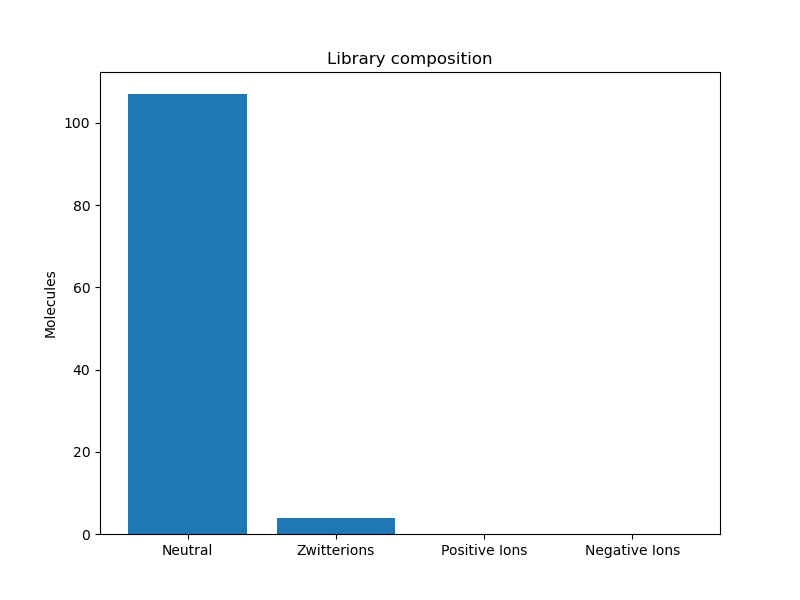}
      \caption{Molecule types in the Literature test set.}
       \label{fig:test_literature_libcomp}
\end{figure}

\begin{figure}[H]
\centering
     \includegraphics[width=1.0\textwidth]{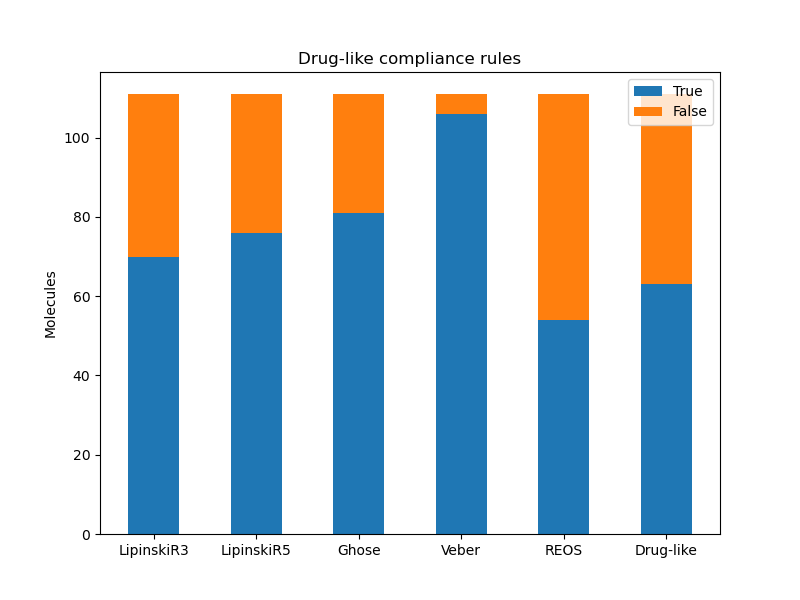}
      \caption{Number of molecules compliant with common drugability rules in the Literature test set.}
       \label{fig:test_literature_compliance}
\end{figure}

\begin{figure}[H]
\centering
     \includegraphics[width=1.0\textwidth]{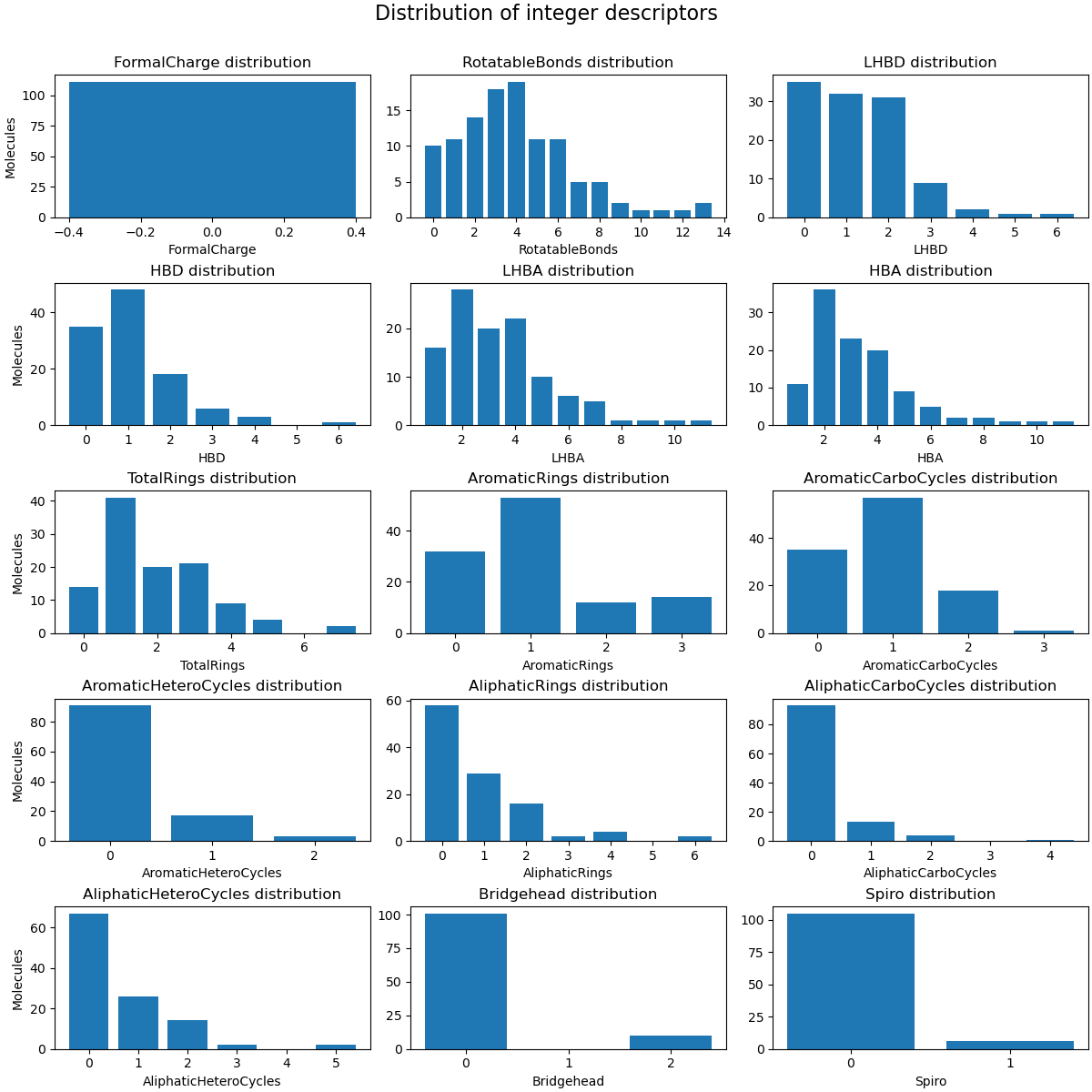}
      \caption{Additional integer dataset descriptors for the Literature test set. Abbreviations: LHBD=Lipinski Hydrogen Bond Donors, HBD=Hydrogen Bond Donors, LHBA=Lipinski Hydrogen Bond Acceptors, HBA=Hydrogen Bond Acceptors}
       \label{fig:test_literature_intdesc}
\end{figure}

\begin{figure}[H]
\centering
     \includegraphics[width=1.0\textwidth]{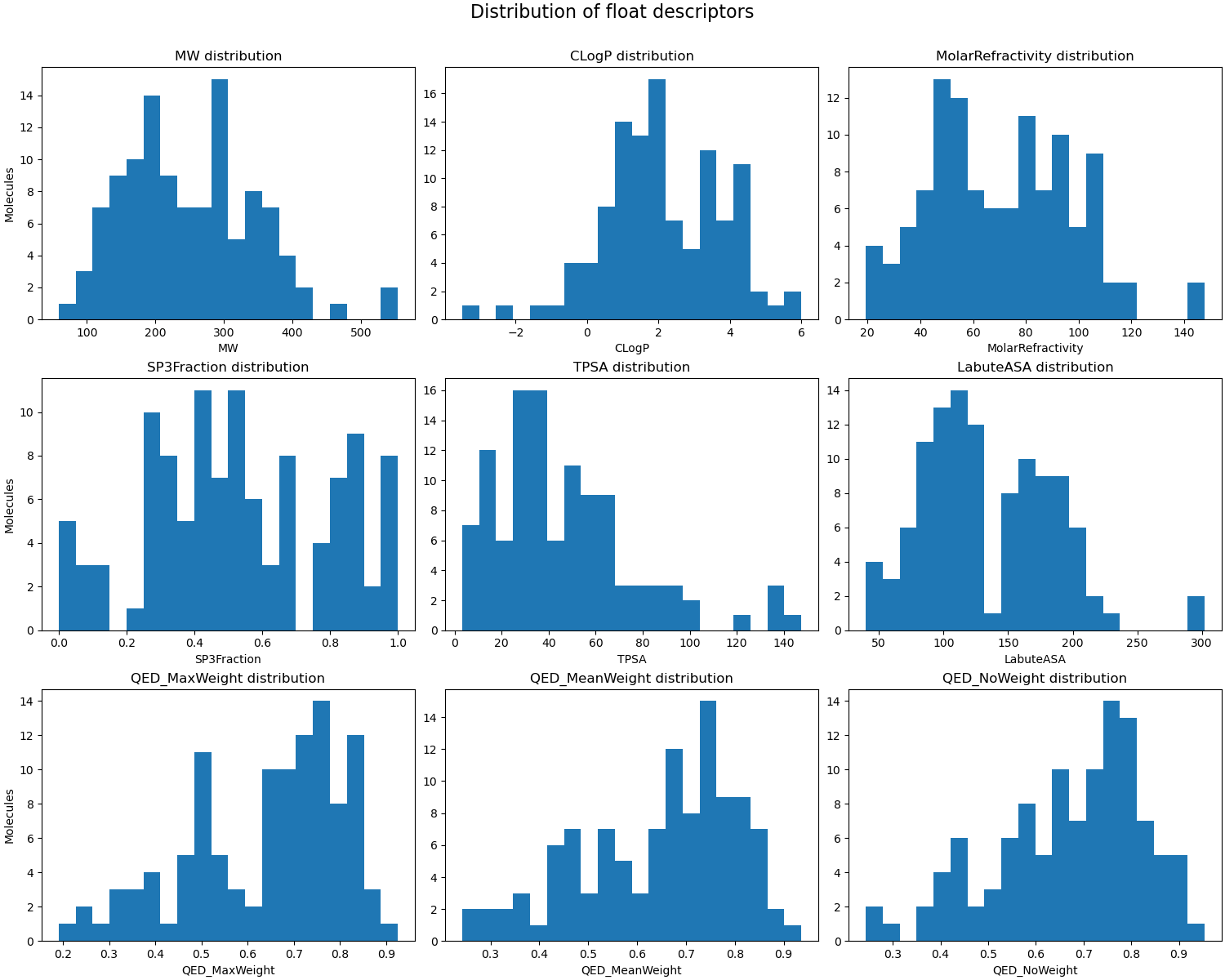}
      \caption{Additional float dataset descriptors for the Literature test set. Abbreviations: TPSA=Topological Polar Surface Area, LabuteASA=Labute Accessible Surface Area, QED=Quantitative Estimate of Druglikeness}
       \label{fig:test_literature_floatdesc}
\end{figure}

\subsubsection{Novartis Test Set Additional Statistics}

\begin{figure}[H]
\centering
     \includegraphics[width=1.0\textwidth]{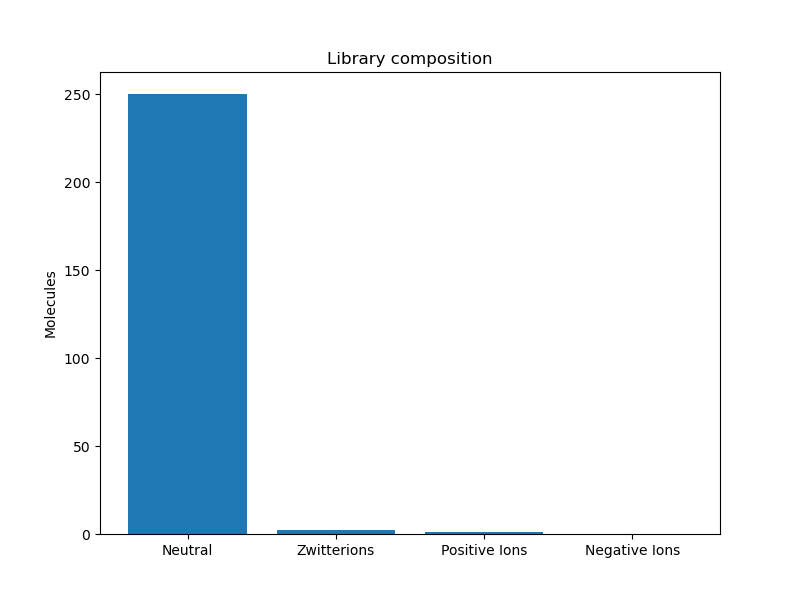}
      \caption{Molecule types in the Novartis test set.}
       \label{fig:test_novartis_libcomp}
\end{figure}

\begin{figure}[H]
\centering
     \includegraphics[width=1.0\textwidth]{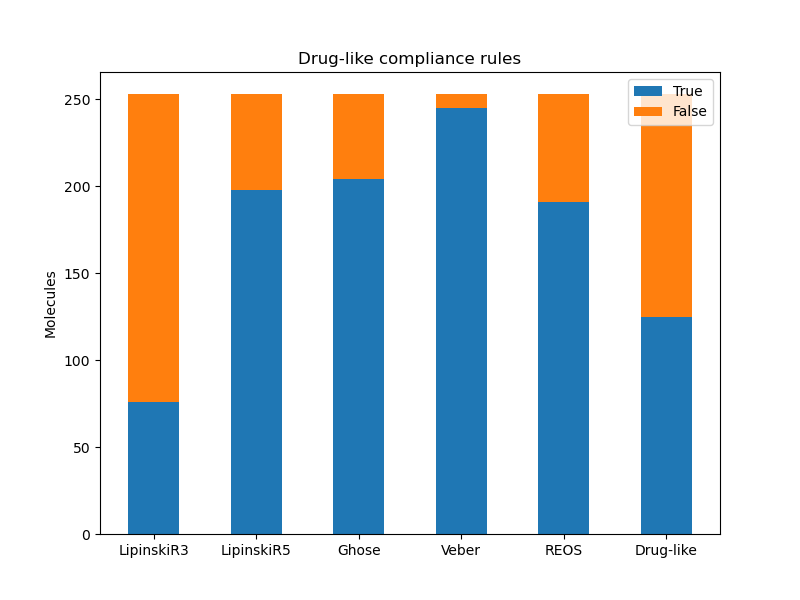}
      \caption{Number of molecules compliant with common drugability rules in the Novartis test set.}
       \label{fig:test_novartis_compliance}
\end{figure}

\begin{figure}[H]
\centering
     \includegraphics[width=1.0\textwidth]{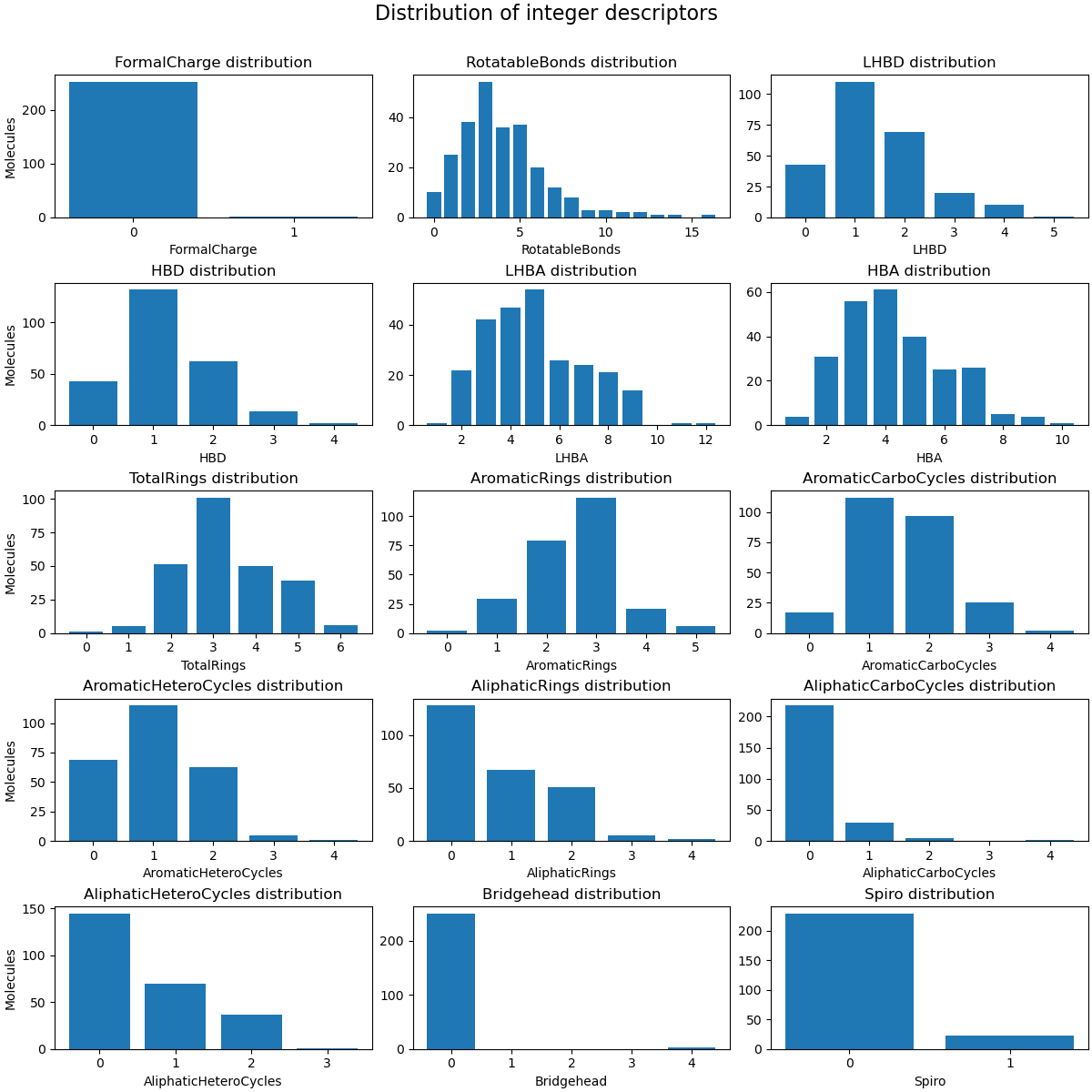}
      \caption{Additional integer dataset descriptors for the Novartis test set. Abbreviations: LHBD=Lipinski Hydrogen Bond Donors, HBD=Hydrogen Bond Donors, LHBA=Lipinski Hydrogen Bond Acceptors, HBA=Hydrogen Bond Acceptors}
       \label{fig:test_novartis_intdesc}
\end{figure}

\begin{figure}[H]
\centering
     \includegraphics[width=1.0\textwidth]{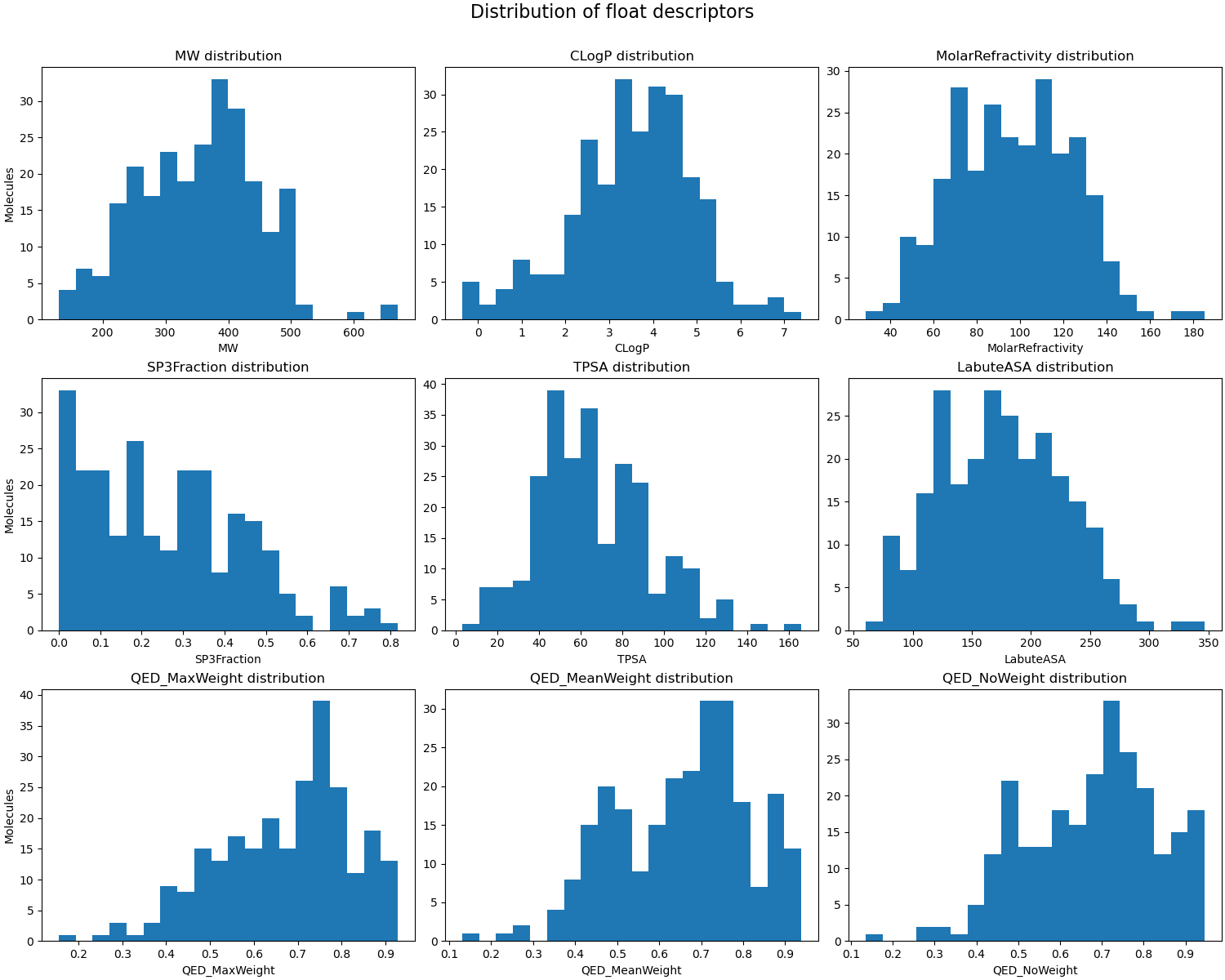}
      \caption{Additional float dataset descriptors for the Novartis test set. Abbreviations: TPSA=Topological Polar Surface Area, LabuteASA=Labute Accessible Surface Area, QED=Quantitative Estimate of Druglikeness}
       \label{fig:test_novartis_floatdesc}
\end{figure}

\subsubsection{Oxy-acids-n-bases Test Set Additional Statistics}

\begin{figure}[H]
\centering
     \includegraphics[width=1.0\textwidth]{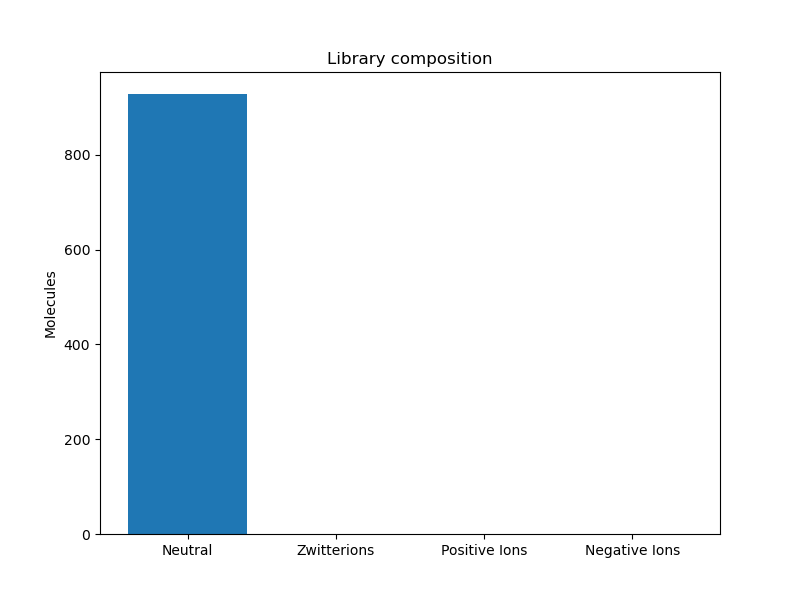}
      \caption{Molecule types in the Oxy-acids-n-bases test set.}
       \label{fig:test_oxy_libcomp}
\end{figure}

\begin{figure}[H]
\centering
     \includegraphics[width=1.0\textwidth]{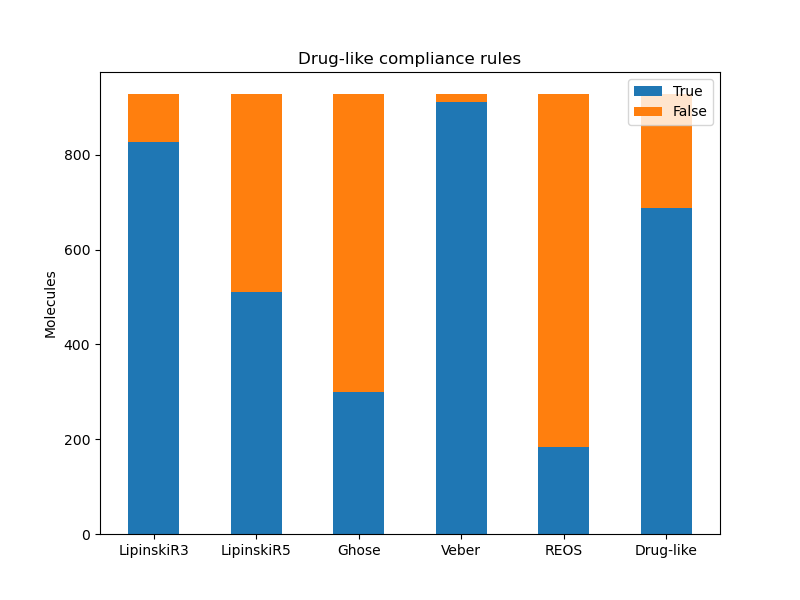}
      \caption{Number of molecules compliant with common drugability rules in the Oxy-acids-n-bases test set.}
       \label{fig:test_oxy_compliance}
\end{figure}

\begin{figure}[H]
\centering
     \includegraphics[width=1.0\textwidth]{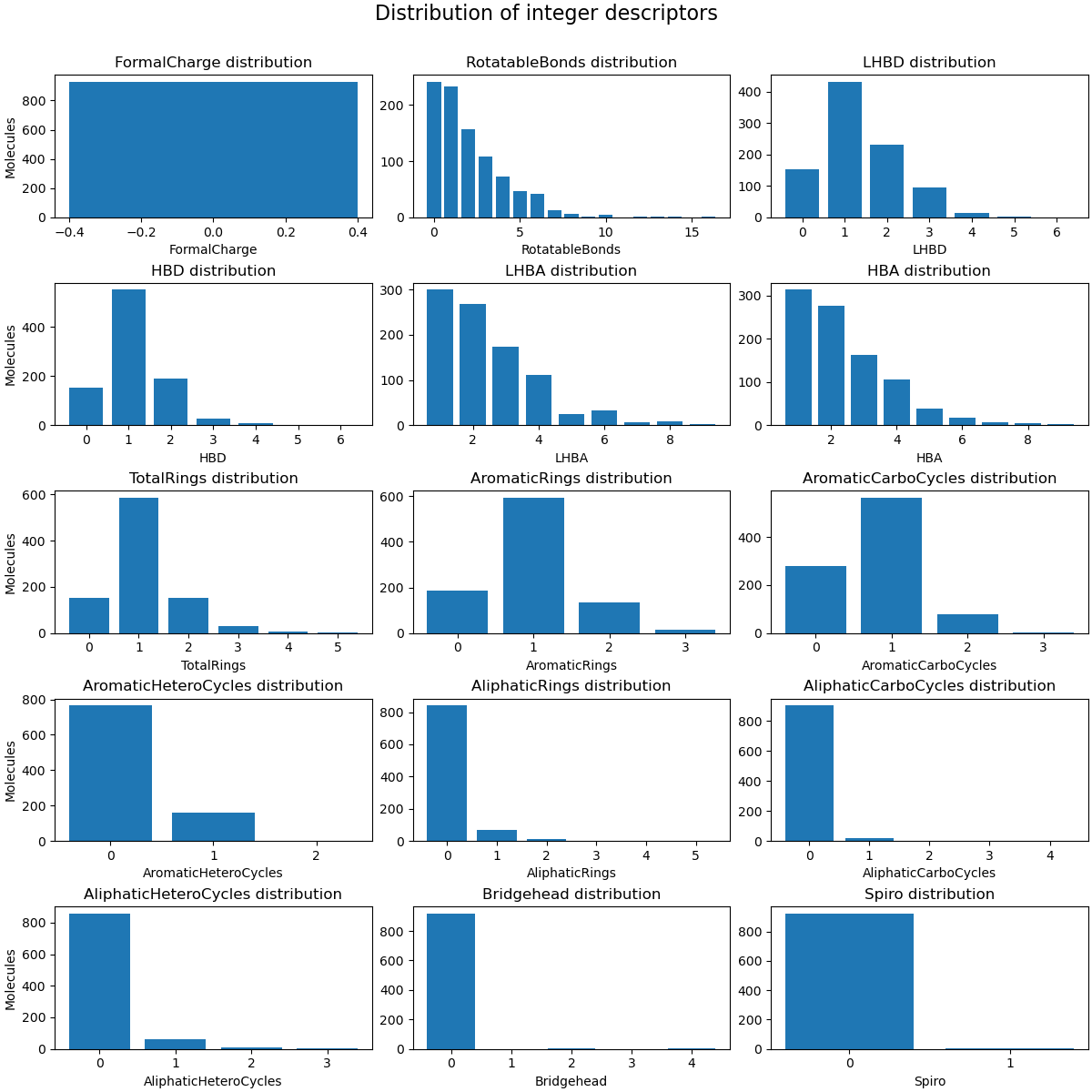}
      \caption{Additional integer dataset descriptors for the Oxy-acids-n-bases test set. Abbreviations: LHBD=Lipinski Hydrogen Bond Donors, HBD=Hydrogen Bond Donors, LHBA=Lipinski Hydrogen Bond Acceptors, HBA=Hydrogen Bond Acceptors}
       \label{fig:test_oxy_intdesc}
\end{figure}

\begin{figure}[H]
\centering
     \includegraphics[width=1.0\textwidth]{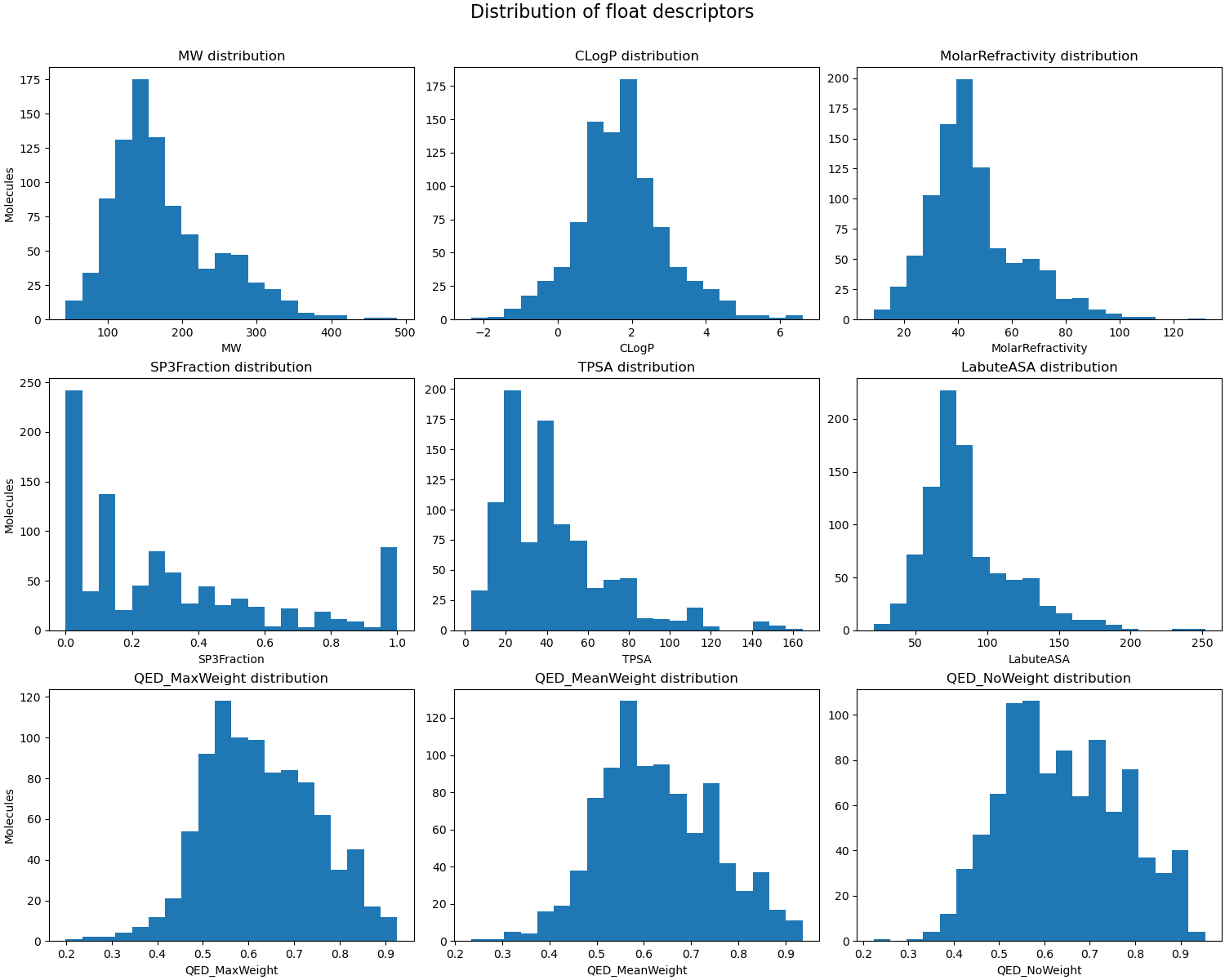}
      \caption{Additional float dataset descriptors for the Oxy-acids-n-bases test set. Abbreviations: TPSA=Topological Polar Surface Area, LabuteASA=Labute Accessible Surface Area, QED=Quantitative Estimate of Druglikeness}
       \label{fig:test_oxy_floatdesc}
\end{figure}

\subsubsection{Transformations Test Set Additional Statistics}

\begin{figure}[H]
\centering
     \includegraphics[width=1.0\textwidth]{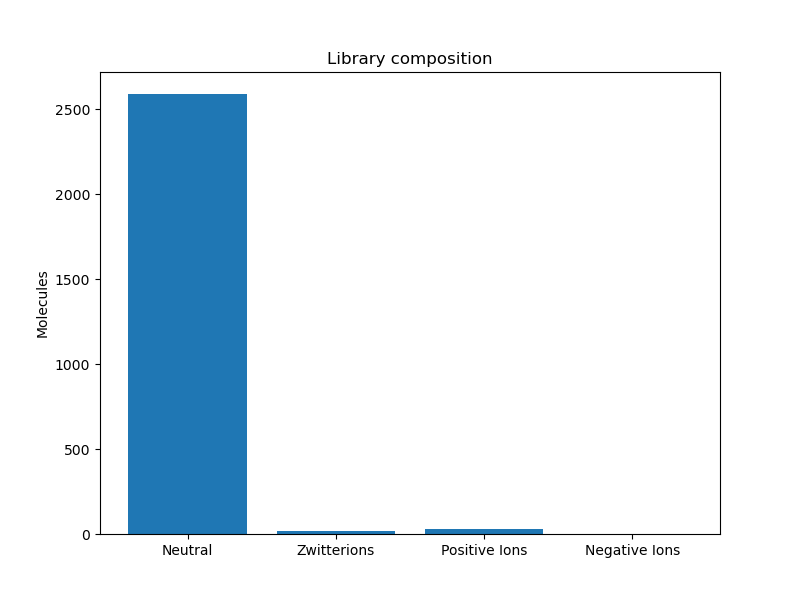}
      \caption{Molecule types in the Transformations test set.}
       \label{fig:test_transform_libcomp}
\end{figure}

\begin{figure}[H]
\centering
     \includegraphics[width=1.0\textwidth]{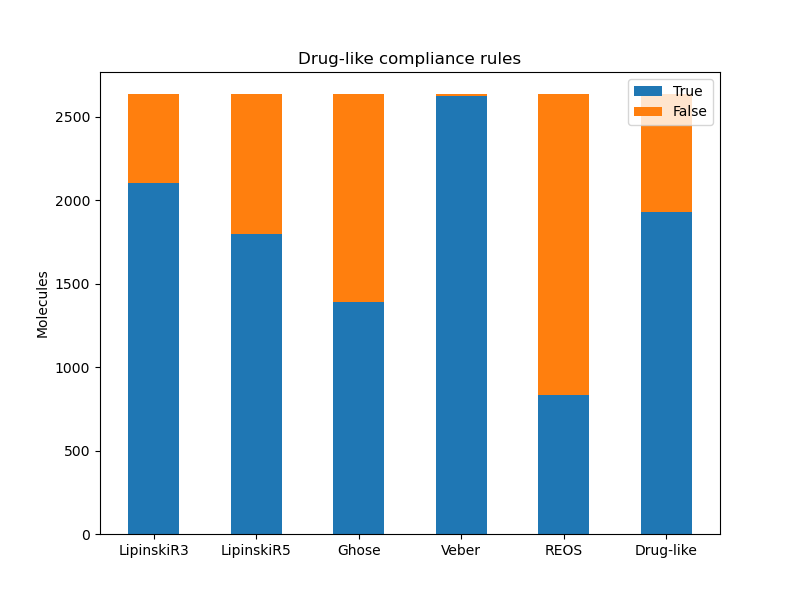}
      \caption{Number of molecules compliant with common drugability rules in the Transformations test set.}
       \label{fig:test_transform_compliance}
\end{figure}

\begin{figure}[H]
\centering
     \includegraphics[width=1.0\textwidth]{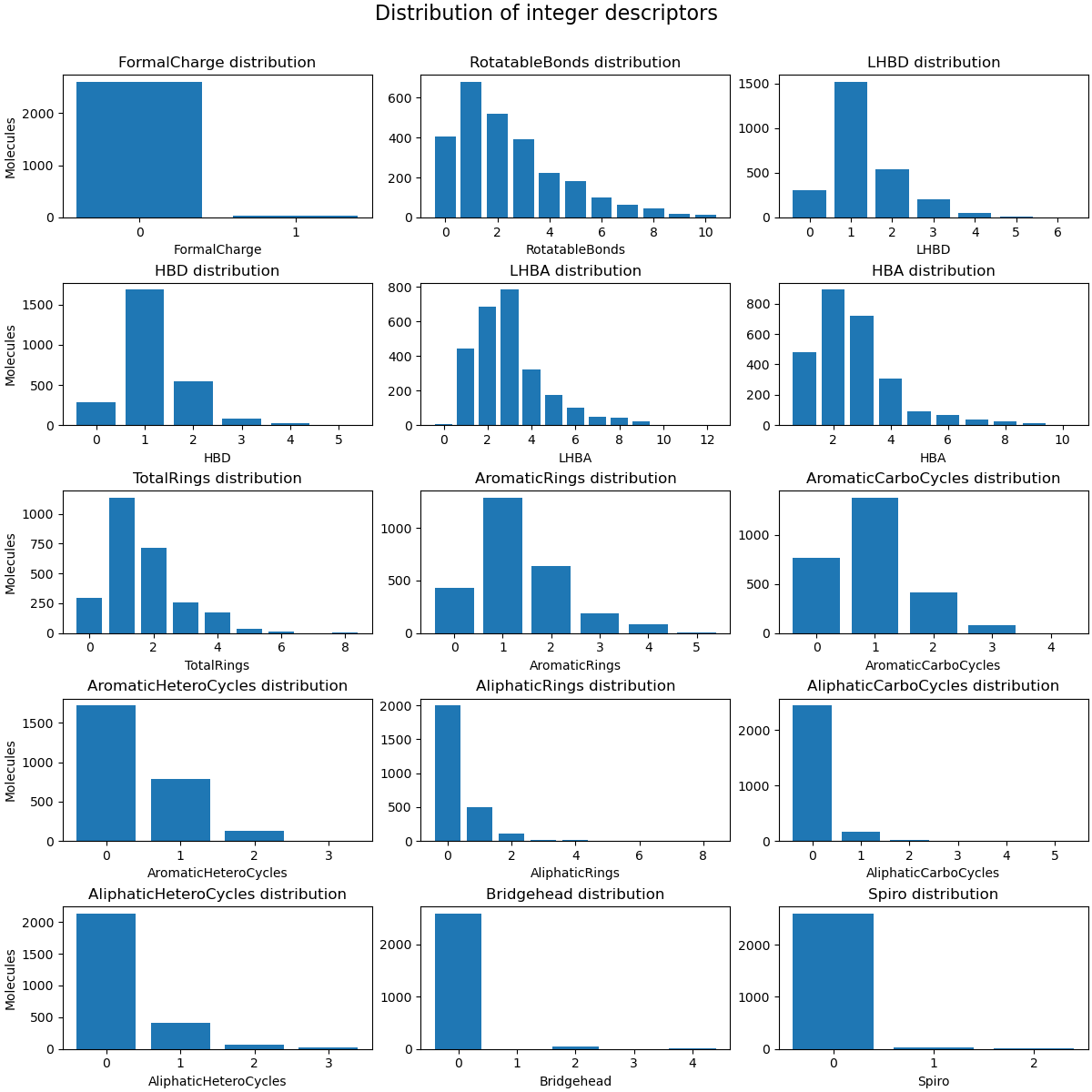}
      \caption{Additional integer dataset descriptors for the Transformations test set. Abbreviations: LHBD=Lipinski Hydrogen Bond Donors, HBD=Hydrogen Bond Donors, LHBA=Lipinski Hydrogen Bond Acceptors, HBA=Hydrogen Bond Acceptors}
       \label{fig:test_transform_intdesc}
\end{figure}

\begin{figure}[H]
\centering
     \includegraphics[width=1.0\textwidth]{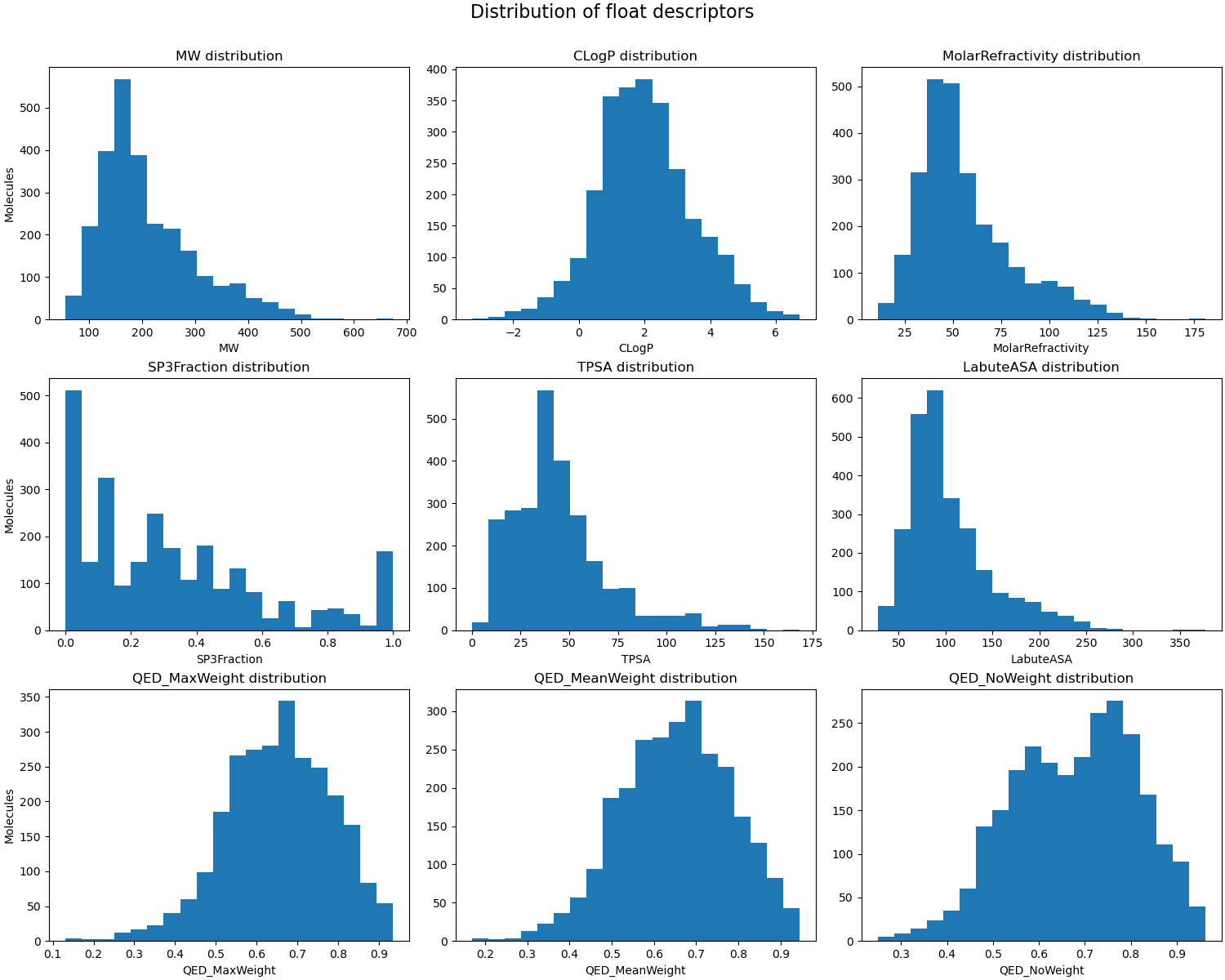}
      \caption{Additional float dataset descriptors for the Transformations test set. Abbreviations: TPSA=Topological Polar Surface Area, LabuteASA=Labute Accessible Surface Area, QED=Quantitative Estimate of Druglikeness}
       \label{fig:test_transform_floatdesc}
\end{figure}

\section{Training Sets Performance Comparison}

\begin{figure}[H]
\centering
     \includegraphics[width=1.0\textwidth]{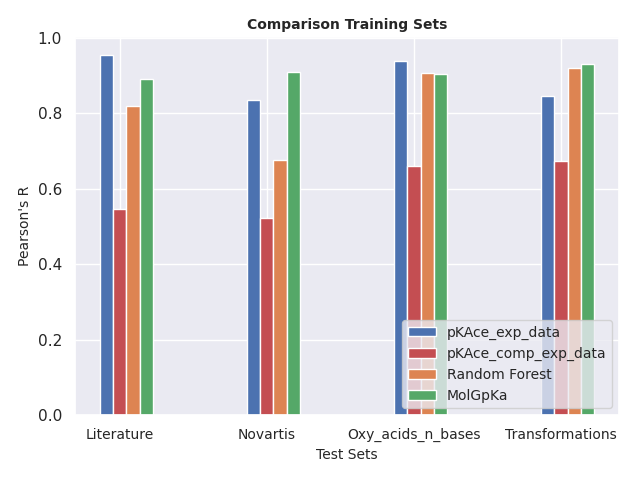}
      \caption{Benchmark results of pKAce model trained on different training set compositions of experimental and computed values and tested against reference models on the 4 external benchmark test sets.}
       \label{fig:training_set_comparison}
\end{figure}

\section{Benchmark Results}

\begin{figure}[H]
\centering
     \includegraphics[width=1.0\textwidth]{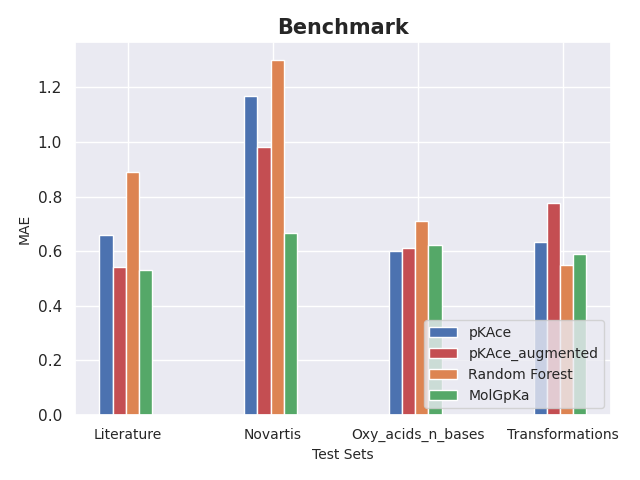}
      \caption{Benchmark results of pKAce model trained on augmented and non-augmented training datasets and tested against reference models on the 4 external benchmark test sets. Mean absolute error values.}
       \label{fig:augmented_benchmark_mae}
\end{figure}

\section{Multiple Conformers Test}
\begin{table}[H]
\resizebox{\textwidth}{!}{%
\begin{tabular}{l|ll}
                          & \multicolumn{2}{c}{Predicted Value} \\ 
 &
  \multicolumn{1}{c}{\begin{tabular}[c]{@{}c@{}}Model trained with\\ multiple conformers\end{tabular}} &
  \multicolumn{1}{c}{\begin{tabular}[c]{@{}c@{}}Model trained with\\ single conformer\end{tabular}} \\ \cline{2-3} 
Conformer 1               & 8.65             & 8.39             \\
Conformer 2               & 8.82             & 8.43             \\
Conformer 3               & 8.62             & 8.61             \\
Conformer 4               & 8.82             & 8.29             \\
Conformer 5               & 8.84             & 8.28             \\
Conformer 6               & 8.44             & 8.37             \\
Conformer 7               & 8.81             & 8.50             \\
Conformer 8               & 8.84             & 8.97             \\ \hline
Stdev. across predictions & \textbf{0.15}    & \textbf{0.31}    \\
\begin{tabular}[c]{@{}l@{}}Mean abs. error to\\ experimental value\end{tabular} &
  \textbf{0.12} &
  \textbf{0.42}
\end{tabular}%
}
\caption{Comparison of standard deviation in predictions for a random test molecule from the Literature test set generated with a pKAce model trained on multiple conformers versus a model trained on single conformers only. Provided is also the average mean absolute error between the model's predictions for each conformer and the true experimental pKa value.}
\label{tab:individual_conformer_preds}
\end{table}

\printbibliography